\documentclass[twocolumn]{aastex631}

\usepackage{xspace, color, comment}
\usepackage{amsmath, amssymb, booktabs}
\newcommand{\name}{SN~2023ufx\xspace}

\newcommand{\Rstar}{\hbox{\ensuremath{R_{\star}}}\xspace}
\newcommand{\Rsun}{\hbox{\ensuremath{\rm R_\odot}}\xspace}

\newcommand{\Msun}{\hbox{\ensuremath{\rm M_\odot}}\xspace}
\newcommand{\kms}{\hbox{\ensuremath{\rm{km}~\rm s^{-1}}}\xspace}

\newcommand{\Menv}{\hbox{\ensuremath{M_{\rm env}}}\xspace}

\newcommand{\texp}{\hbox{\ensuremath{t_{\rm exp}}}\xspace}

\newcommand{\Mni}{\hbox{\ensuremath{M_{\rm Ni}}}\xspace}
\newcommand{\Mdot}{\hbox{\ensuremath{\dot{M}}}\xspace}

\newcommand{\sneii}{SNe~II\xspace}
\newcommand{\snii}{SN~II\xspace}
\newcommand{\Ha}{H$\alpha$\xspace}
\newcommand{\Hb}{H$\beta$\xspace}

\newcommand{\uJy}{\hbox{\ensuremath{\mu\rm{Jy}}}\xspace}
\newcommand{\fnu}{\hbox{\ensuremath{F_\nu}}\xspace}

\newcommand{\swift}{\emph{Swift}\xspace}

\newcommand{\AAA}{\hbox{\ensuremath{\rm\AA}}\xspace}
\newcommand{\um}{\hbox{\ensuremath{\mu\rm m}}\xspace}

\newcommand{\Pa}{P$\alpha$\xspace}
\newcommand{\Pb}{P$\beta$\xspace}
\newcommand{\Pg}{P$\gamma$\xspace}
\newcommand{\Hg}{H$\gamma$\xspace}
\newcommand{\Hd}{H$\delta$\xspace}

\newcommand{\sneiib}{SNe~IIb\xspace}

\newcommand{\OIIlong}{[\ion{O}{2}]$\lambda3727$\xspace}
\newcommand{\OII}{[\ion{O}{2}]\xspace}
\newcommand{\OIII}{[\ion{O}{3}]\xspace}

\newcommand{\OI}{[\ion{O}{1}]\xspace}
\newcommand{\OIlong}{[\ion{O}{1}]$\lambda\lambda6300,6363$\xspace}
\newcommand{\OIa}{[\ion{O}{1}]$\lambda6300$\xspace}

\newcommand{\OIc}{[\ion{O}{1}]$\lambda5577$\xspace}
\newcommand{\sne}{SNe\xspace}

\newcommand{\NIIlong}{\hbox{[\ion{N}{2}]$\lambda6584$}\xspace}
\newcommand{\SII}{[\ion{S}{2}]\xspace}

\newcommand{\FeII}{\ion{Fe}{2}\xspace}
\newcommand{\FeIIa}{\ion{Fe}{2}$\lambda5169$\xspace}

\newcommand{\ccsne}{CC~SNe\xspace}

\defcitealias{scat-spec}{269}
\defcitealias{disc}{268}
\defcitealias{class}{278}
\defcitealias{redshift}{292}

\newcommand{\dtpl}{\hbox{\ensuremath{\Delta t_{\rm pl}}}\xspace}

\shorttitle{The Metal-Poor SN II 2023ufx}
\shortauthors{M. A. Tucker et al.}
\begin{document}

\title{The Extremely Metal-Poor SN~2023ufx: A Local Analog to High-Redshift Type II Supernovae}

\author[0000-0002-2471-8442]{Michael A. Tucker}
\altaffiliation{CCAPP Fellow}
\affiliation{Center for Cosmology and AstroParticale Physics, 191 W Woodruff Ave, Columbus, OH 43210}
\affiliation{Department of Astronomy, The Ohio State University, 140 W 18th Ave, Columbus, OH 43210}

\correspondingauthor{Michael Tucker}
\email{tuckerma95@gmail.com}

\author[0000-0001-9668-2920]{Jason Hinkle}
\altaffiliation{NASA FINESST Future Investigator}
\affiliation{Institute for Astronomy, University of Hawai`i, 2680 Woodlawn Drive, Honolulu HI 96822}

\author[0000-0002-4269-7999]{Charlotte R. Angus}
\affiliation{Astrophysics Research Centre, School of Mathematics and Physics, Queen's University Belfast, Belfast BT7 1NN, UK}

\author[0000-0002-4449-9152]{Katie Auchettl}
\affiliation{School of Physics, The University of Melbourne, VIC 3010, Australia}
\affiliation{Department of Astronomy and Astrophysics, University of California, Santa Cruz, CA 95064, USA}

\author[0000-0003-3953-9532]{Willem~B.~Hoogendam}
\altaffiliation{NSF Graduate Research Fellow}
\affiliation{Institute for Astronomy, University of Hawai`i, 2680 Woodlawn Drive, Honolulu HI 96822}

\author[0000-0003-4631-1149]{Benjamin Shappee}
\affiliation{Institute for Astronomy, University of Hawai`i, 2680 Woodlawn Drive, Honolulu HI 96822}

\author[0000-0001-6017-2961]{Christopher S. Kochanek}
\affiliation{Center for Cosmology and AstroParticale Physics, 191 W Woodruff Ave, Columbus, OH 43210}
\affiliation{Department of Astronomy, The Ohio State University, 140 W 18th Ave, Columbus, OH 43210}

\author[0000-0002-5221-7557]{Chris Ashall}
\affiliation{Department of Physics, Virginia Tech, 850 West Campus Drive, Blacksburg VA, 24061, USA}
\affiliation{Institute for Astronomy, University of Hawai`i, 2680 Woodlawn Drive, Honolulu HI 96822}

\author{Thomas de Boer}
\affiliation{Institute for Astronomy, University of Hawai`i, 2680 Woodlawn Drive, Honolulu HI 96822}

\author{Kenneth C. Chambers}
\affiliation{Institute for Astronomy, University of Hawai`i, 2680 Woodlawn Drive, Honolulu HI 96822}

\author[0000-0002-2164-859X]{Dhvanil D. Desai}
\affiliation{Institute for Astronomy, University of Hawai`i, 2680 Woodlawn Drive, Honolulu HI 96822}

\author[0000-0003-3429-7845]{Aaron Do}
\affiliation{Institute of Astronomy and Kavli Institute for Cosmology, Madingley Road, Cambridge, CB3 0HA, UK}

\author[0000-0003-1916-0664]{Michael D. Fulton}
\affiliation{Astrophysics Research Centre, School of Mathematics and Physics, Queen's University Belfast, Belfast BT7 1NN, UK}

\author[0000-0003-1015-5367]{Hua Gao}
\affiliation{Institute for Astronomy, University of Hawai`i, 2680 Woodlawn Drive, Honolulu HI 96822}

\author{Joanna Herman}
\affiliation{Institute for Astronomy, University of Hawai`i, 2680 Woodlawn Drive, Honolulu HI 96822}

\author[0000-0003-1059-9603]{Mark Huber}
\affiliation{Institute for Astronomy, University of Hawai`i, 2680 Woodlawn Drive, Honolulu HI 96822}

\author{Chris Lidman}
\affiliation{The Research School of Astronomy and Astrophysics, The Australian National University, ACT 2601, Australia}

\author[0000-0002-7272-5129]{Chien-Cheng Lin}
\affiliation{Institute for Astronomy, University of Hawai`i, 2680 Woodlawn Drive, Honolulu HI 96822}

\author{Thomas B. Lowe}
\affiliation{Institute for Astronomy, University of Hawai`i, 2680 Woodlawn Drive, Honolulu HI 96822}

\author[0000-0002-7965-2815]{Eugene A. Magnier}
\affiliation{Institute for Astronomy, University of Hawai`i, 2680 Woodlawn Drive, Honolulu HI 96822}

\author[0009-0006-4963-3206]{Bailey Martin}
\affiliation{The Research School of Astronomy and Astrophysics, The Australian National University, ACT 2601, Australia}

\author{Paloma Mínguez}
\affiliation{Institute for Astronomy, University of Hawai`i, 2680 Woodlawn Drive, Honolulu HI 96822}

\author[0000-0002-2555-3192]{Matt Nicholl}
\affiliation{Astrophysics Research Centre, School of Mathematics and Physics, Queen's University Belfast, Belfast BT7 1NN, UK}

\author[0000-0003-4663-4300]{Miika Pursiainen}
\affiliation{Department of Physics, University of Warwick, Gibbet Hill Road, Coventry, CV4 7AL, UK}

\author[0000-0002-8229-1731]{S. J. Smartt}
\affiliation{Astrophysics sub-Department, Department of Physics, University of Oxford, Keble Road, Oxford, OX1 3RH, UK}
\affiliation{Astrophysics Research Centre, School of Mathematics and Physics, Queen's University Belfast, Belfast BT7 1NN, UK}

\author[0000-0001-9535-3199]{Ken W. Smith}
\affiliation{Astrophysics Research Centre, School of Mathematics and Physics, Queen's University Belfast, Belfast BT7 1NN, UK}

\author[0000-0003-4524-6883]{Shubham Srivastav}
\affiliation{Astrophysics sub-Department, Department of Physics, University of Oxford, Keble Road, Oxford, OX1 3RH, UK}

\author[0000-0002-4283-5159]{Brad~E.~Tucker}
\affiliation{Mt Stromlo Observatory, The Research School of Astronomy and Astrophysics, Australian National University, ACT 2611, Australia}
\affiliation{National Centre for the Public Awareness of Science, Australian National University, ACT 2601, Australia}
\affiliation{The ARC Centre of Excellence for All-Sky Astrophysics in 3 Dimension (ASTRO 3D), Australia}

\author[0000-0002-1341-0952]{Richard~J.~Wainscoat}
\affiliation{Institute for Astronomy, University of Hawai`i, 2680 Woodlawn Drive, Honolulu HI 96822}

%%%%%%%%%%%%%%%%%%%%%%%%%%%%
%%% place new author info below this line 
%-----------------------------------------



\begin{abstract}
We present extensive observations of the Type II supernova (SN~II) 2023ufx which is likely the most metal-poor \snii observed to-date. It exploded in the outskirts of a low-metallicity ($Z_{\rm host} \sim 0.1~Z_\odot$) dwarf ($M_g = -13.39\pm0.16$~mag; $r_{\rm proj}\sim 1$~kpc) galaxy. The explosion is luminous, peaking at $M_g\approx -18.5$~mag, and shows rapid evolution. The $r$-band (pseudo-bolometric) light curve has a shock-cooling phase lasting 20 (17) days followed by a 19 (23)-day plateau. The entire optically-thick phase lasts only $\approx 55$~days following explosion, indicating that the red supergiant progenitor had a thinned H envelope prior to explosion. The early spectra obtained during the shock-cooling phase show no evidence for narrow emission features and limit the pre-explosion mass-loss rate to $\Mdot \lesssim 10^{-3}$~\Msun/yr. The photospheric-phase spectra are devoid of prominent metal absorption features, indicating a progenitor metallicity of $\lesssim 0.1~Z_\odot$. The semi-nebular ($\sim 60-130$~d) spectra reveal weak \ion{Fe}{2}, but other metal species typically observed at these phases (\ion{Ti}{2}, \ion{Sc}{2}, \ion{Ba}{2}) are conspicuously absent. The late-phase optical and near-infrared spectra also reveal broad ($\approx10^4~\kms$) double-peaked \Ha, \Pb, and \Pg emission profiles suggestive of a fast outflow launched during the explosion. Outflows are typically attributed to rapidly-rotating progenitors which also prefer metal-poor environments. This is only the second SN~II with $\lesssim 0.1~Z_\odot$ and both exhibit peculiar evolution, suggesting a sizable fraction of metal-poor SNe~II have distinct properties compared to nearby metal-enriched SNe~II. These observations lay the groundwork for modeling the metal-poor \sneii expected in the early Universe.

\end{abstract}



\section{Introduction} \label{sec:intro}

H-rich Type II supernovae (SNe~II) originate from the core-collapse of red supergiants (RSGs, \citealp{smartt2009a, smartt2009b}). Observations within the first $\sim$week probe the cooling of the shock-heated envelope (the shock-cooling phase; e.g., \citealp{sapir2017}) before the ejecta transitions to the plateau phase where recombination of H in the shock-ionized envelope produces a quasi-constant optical luminosity \citep{falk1977}. Plateau-phase spectra are characterized by prominent P-Cygni Balmer features and absorption from various metal species such as Fe, Sc, and Ba \citep[e.g., ][]{gutierrez2017a}. The light curve transitions to the radioactive decay tail after the recombination wave reaches the He zone where the observed emission is powered by the radioactive decay of $^{56}$Co. Nebular spectra constrain the mass of the progenitor star through the strength of emission lines from intermediate-mass elements (IMEs; i.e., O, Ca; e.g., \citealp{jerkstrand2012, jerkstrand2014}). Calibrated to progenitor stars identified in pre-explosion imaging \citep[e.g., ][]{li2005, mattila2008, maund2014, kochanek2017, kilpatrick2018}, there is growing agreement between \snii models and observations \citep[e.g., ][]{dessart2013, dessart2017, utrobin2017b, martinez2019}.

The duration of the plateau phase (\dtpl) primarily depends on the mass of the H envelope (\Menv) with minor contributions from the explosion energy and synthesized $^{56}$Ni mass (\Mni; \citealp{popov1993, kasen2009}). \sneii are observationally classified according to their plateau durations. On the long end of the \dtpl distribution are the canonical SNe~IIP exhibiting $\dtpl \sim 80-120$~days \citep[e.g., ][]{barker2022}. SNe~IIb, with negligible H envelopes, occupy the short end of the \dtpl distribution ($\dtpl\sim 0$~days; \citealp{chevalier2010, gilkis2022}) and the intermediate short-plateau and IIL SNe have thin H envelopes (e.g., \citealp{hiramatsu2021}). Theoretically, decreasing plateau durations correspond to lower \Menv. SNe II with $\dtpl \lesssim 80$~days suggest the RSG had a thinner H envelope than expected from a typical $10-20~\Msun$ \snii progenitor. This can be interpreted as a mass sequence of \snii progenitors where more massive stars have stronger stellar winds that more efficiently remove their envelopes \citep[e.g., ][]{dejager1988, vink2001, vink2022} or due to interaction with a nearby binary companion \citep[e.g., ][]{eldridge2018, dessart2024}. Disentangling these effects remains a major difficulty in understanding massive star formation and (co-)evolution \citep[e.g., ][]{langer2012, smith2014}.

Progenitor metallicity ($Z$) is another poorly-constrained aspect of massive-star evolution \citep[e.g., ][]{sanyal2017, chun2018, volpato2023} and their SNe \citep{ibeling2013, dessart2013, dessart2014}. For a given zero-age main sequence (ZAMS) mass, reduced metallicity lowers mass-loss rates with $\Mdot\propto Z^\alpha$ and $\alpha \sim 0.3-0.8$ \citep[e.g., ][]{vink2001, mokiem2007, sander2020, vink2022} leading to more massive and compact progenitors \citep{sukhbold2014, limongi2018, aryan2023}. This has implications for stellar feedback \citep[e.g., ][]{ou2023, jecmen2023}, ionizing photon production \citep[e.g., ][]{gotberg2017}, and nucleosynthesis \citep[e.g., ][]{heger2010, limongi2018}. The influence of $Z$ on massive-star evolution is supported by observed correlations between host-galaxy/environmental metallicity and \snii spectral features \citep{prieto2008, dessart2014, galbany2016, anderson2016, taddia2016, pessi2023c}. Low-metallicity and/or dwarf-galaxy \sneii are typically more luminous than the broader population \citep[e.g., ][]{arcavi2010, gutierrez2018, scott2019}. Yet these observed correlations are mostly confined to ensemble analyses of heterogeneous observations that are skewed towards higher-metallicity environments. This is simply a reflection of bright \sne exploding in luminous nearby galaxies where the average stellar population has near-Solar composition. This observational bias also likely accounts for the dearth of $Z \lesssim 0.3~Z_\odot$ \sneii in the samples of \citet{dessart2014} and \citet{anderson2016} which relied on observational campaigns targeting nearby galaxies. Even the untargeted iPTF sample of \citet{taddia2016} only contained a handful ($\sim 5$ out of 39) of candidate low-metallicity \sneii. SN~2015bs \citep{anderson2018} is the only known \snii with a reliable metallicity $\lesssim 0.1~Z_\odot$ but lacks extensive spectroscopic observations. As we push deeper into the early Universe with JWST \citep[e.g., ][]{topping2022, boyett2022}, a holistic understanding of metal-poor massive stars, their SNe and subsequent feedback is paramount.

Here we present follow-up observations and analysis of the luminous, fast-evolving, and metal-poor \snii \name. The observations and data reductions procedures are summarized in \S\ref{sec:data}. We analyze the host-galaxy properties in \S\ref{sec:host}, finding that \name exploded in the outskirts of a metal-poor dwarf galaxy. We analyze the photometric and spectroscopic observations of \name in \S\ref{sec:phot} and \S\ref{sec:spex}, respectively. We place \name in broader context and discuss its implications in \S\ref{sec:discuss}. We review our findings in \S\ref{sec:summary}. A full accounting of the data and calibration are given in Appendix~\ref{app:obs}. Additional figures and tables are included in Appendix~\ref{app:extra} along with the references for comparison SNe. The main properties of \name and its host are summarized in Table~\ref{tab:hosttbl}.

\section{Observations}\label{sec:data}
\name was discovered by the Asteroid Terrestial-impact Last Alert System \citep[ATLAS; ][]{tonry2018a, smithATLAS20} on UT 2023-10-06 13:55:52 (MJD 60223.58) at $o\approx18.8$~mag (internal designation ATLAS23tsa). We summarize the photometry and spectroscopy here and provide a detailed accounting of the data reduction and calibration in Appendix~\ref{app:obs}. All observations are corrected for Milky Way reddening of $E(B-V)=0.04$ \citep{schlafly2011} using $R_V=3.1$ and the wavelength dependence of \citet{fitzpatrick1999}. No correction for host-galaxy extinction is applied because it is negligible based on the consistency of the early spectra with an un-reddened blackbody, the lack of \ion{Na}{1} absorption at the host redshift, and the distance of \name from the host.

We collect survey light curves including $c$- and $o$-band photometry from ATLAS, $g$-band photometry from the All-Sky Automated Survey for SuperNovae \citep[ASAS-SN; ][]{shappee2014}, $g$- and $r$-band photometry from the Zwicky Transient Facility \citep[ZTF; ][]{bellm2019}, and $grizyw$ photometry from the Panoramic Survey Telescope and Rapid Response System \citep[Pan-STARRS; ][]{chambers2016}. Daily average fluxes are computed for surveys with more than one observation per-night but we do not combine across different surveys (e.g., ZTF $g$ and Pan-STARRS $g$) due to small differences in system throughputs. Additional photometric observations were obtained with the \emph{Swift} UltraViolet and Optical Telescope \citep[UVOT; ][]{gehrels2004, roming2005} and the Multi-Object Double Spectrograph \citep{pogge2010}. The \emph{Swift} observations cover the $UVW2$, $UVM2$, $UVW1$, $U$, $B$, and $V$ filters and the MODS imaging used the $ugri$ filters.

Multiple epochs of optical and near-IR (NIR) spectra were obtained for \name. The earliest spectrum was obtained $\approx 2.5$~d after discovery, or $\approx 3.5$~d relative to the estimated explosion epoch, by the Spectroscopic Classification of Astronomical Transients \citep[SCAT; ][]{tucker2022} survey using the SuperNova Integral Field Spectrograph \citep[SNIFS; ][]{lantz2004} on the University of Hawai`i 2.2m (UH2.2m) telescope. Other sources of optical spectra include the Multi-Object Double Spectrograph \citep[MODS; ][]{pogge2010} on the Large Binocular Telescope (LBT), the Keck Cosmic Web Imager \citep[KCWI; ][]{kcwi_morrissey} on the Keck~II telescope, the Wide-Field Spectrograph \citep[WiFeS; ][]{wifes1, wifes2} on the Australian National University 2.3m (ANU2.3m) telescope, the Gemini Multi-Object Spectrograph \citep[GMOS; ][]{hook2004} on the Gemini-North (GN) telescope, and the Alhambra Faint Object Spectrograph (ALFOSC)\footnote{\url{https://www.not.iac.es/instruments/alfosc/}} on the Nordic Optical Telescope (NOT). Three epochs of near-infrared (NIR) spectra were obtained with SpeX \citep{rayner03} on NASA's Infrared Telescope Facility (IRTF). The full spectroscopic time-series extends until $\approx 130$~d after explosion. Host-galaxy spectra were extracted from the later KCWI 3D $(x,y,\lambda)$ datacubes once \name had faded sufficiently. Complete details about the data reduction and calibration are provided in Appendix~\ref{app:obs}.

\section{Host-Galaxy Properties}\label{sec:host}

\begin{table}
    \centering
    \begin{tabular}{lr}
    \hline\hline
    Parameter & Value \\\hline\hline
    \multicolumn{2}{c}{SN~2023ufx}\\\hline
    R.A. [hms] & 08:24:51.6 \\
    Decl. [dms] & +21:17:43.2 \\
    R.A. [deg] & 126.2148658 \\
    Decl. [deg] & +21.2953198 \\
    Explosion date [MJD] & $60223.0\pm0.5$\\
    Peak $m_g$~[mag] & $15.55\pm 0.02$\\
    Peak $M_g$~[mag] & $-18.54\pm0.15$\\
    $E(B-V)_{\rm MW}$~[mag] & 0.04 \\
    $E(B-V)_{\rm host}$~[mag] & $0^a$\\
    \hline
    \multicolumn{2}{c}{Host galaxy}\\\hline
    Redshift $z$ & $0.0152\pm0.0001$ \\
    Distance [Mpc] & $65.9\pm4.4$\\
    Distance modulus [mag] & $34.09\pm0.15$\\
    $M_g$ [mag] & $-13.39\pm0.16$ \\
    $M_r$ [mag] & $-13.47\pm0.17$ \\
    Observed $r$-band radius [arcsec] & $3.1\pm0.3$ \\
    Intrinsic $r$-band radius [kpc] & $1.0\pm0.1$ \\
    $\log_{10}(\rm{[OIII]}\lambda5007/\rm{H}\beta)$ [dex] & $0.02\pm0.06$ \\
    $\log_{10}(\rm{[OII]}\lambda3727/\rm{H}\beta)$ [dex] & $0.45\pm0.06$ \\
    $\log_{10}(\rm{[NII]}\lambda6584/\rm{H}\alpha)$ [dex] & $< -1.8 (3\sigma)$ \\
    $\log_{10}(\rm{[OI]}\lambda6300/\rm{H}\alpha)$ [dex] & $< -2.0 (3\sigma)$ \\
    $\log_{10}(\rm{[SII]}\lambda6717,6734/\rm{H}\alpha)$ [dex] & $-1.05\pm0.09$ \\    
    $\log_{10}(\rm O/\rm H)+12$~[dex] & $\lesssim8$ \\
    Host $A_V$ [mag] & $0.00_{-0.00}^{+0.09}$ \\
    $\log_{10}$(Age [yr]) & $8.0_{-0.2}^{+0.4}$ \\
    $\log_{10}$(Mass [$M_\odot$]) & $6.4^{+0.1}_{-0.1}$ \\
    $\log_{10}$(SFR [$\rm M_\odot$/yr]) & $-6.6_{-2.8}^{+4.1}$ \\
    \hline\hline
    \end{tabular}
    \caption{Basic properties of \name and its host.\\$^a$Adopted (see \S\ref{sec:host}).}
    \label{tab:hosttbl}
\end{table}
\begin{figure}
    \centering
    \includegraphics[width=\linewidth]{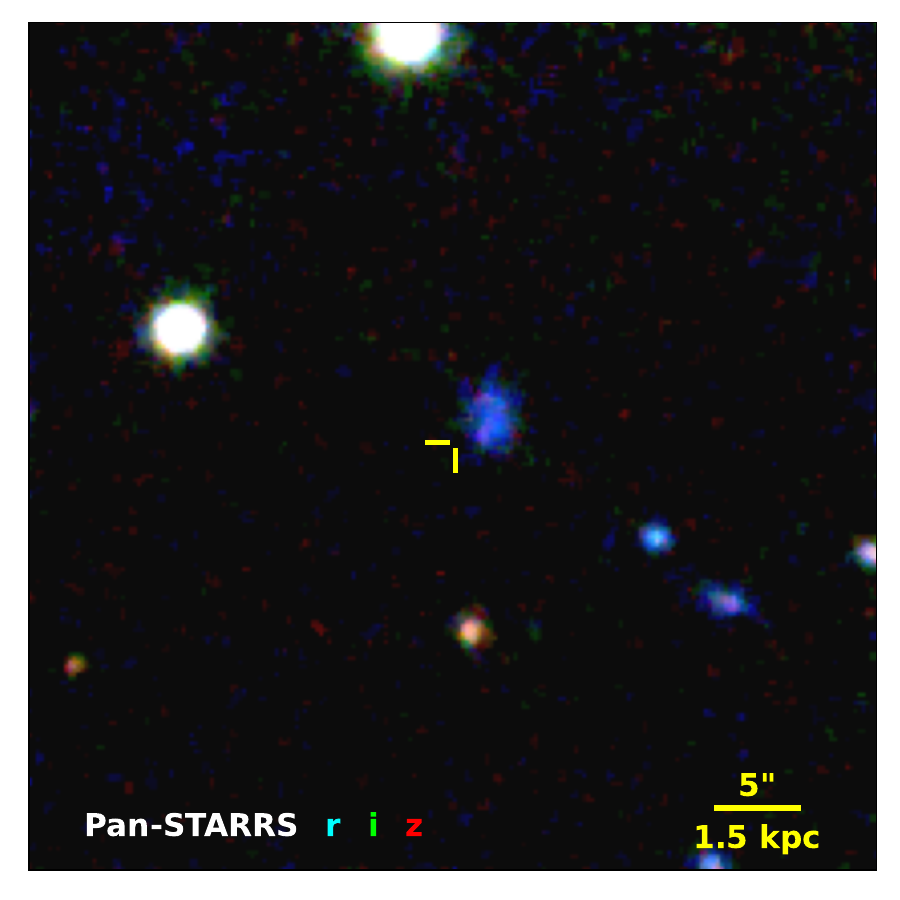}
    \caption{Pre-explosion imaging of the host galaxy from Pan-STARRS. Yellow reticles are 1\arcsec (0.3~kpc) in length and mark the location of \name.}
    \label{fig:cutout}
\end{figure}

Fig.~\ref{fig:cutout} shows the host-galaxy of \name. SDSS J082451.43+211743.3 is a faint ($g = 20.86\pm0.04$~mag) extended blue source with an absolute $g$-band luminosity of $M_g = -13.23 \pm 0.16$~mag. The galaxy is small, with a $g$-band Kron radius \citep{kron1980} of just $R_g = 3\farcs6$ corresponding to a projected intrinsic radius of $1.1$~kpc. \name exploded $\approx 2\farcs7$ ($\sim 0.8$~kpc) from the nucleus. The host-galaxy parameters are summarized in Table~\ref{tab:hosttbl}.

\name was originally considered a potential fast blue optical transient (see AstroNotes \citetalias{disc}, \citetalias{scat-spec}, and \citetalias{class}) based upon the SDSS photometric redshift of $z_{\rm phot} \approx 0.04$. Instead, the true redshift is $z\approx0.015$ (AstroNote \citetalias{redshift}). Fitting the \OII$\lambda3727$, \OIII$\lambda\lambda4959,5007$, \Hb, \Ha, and \SII$\lambda\lambda6716,6731$ host-galaxy emission lines in the late-time KCWI observations produces a refined redshift of $z=0.0152\pm0.0001$. At this redshift, the host-galaxy is located at $d \approx 65.9\pm4.4$~Mpc based on the dynamical velocity-field estimate of \citet{shaya2022}, corresponding to a distance modulus of $34.09\pm0.15$~mag using cosmological parameters $H_0 = 70$~km/s/Mpc and $\Omega_m = 0.3$.

The host galaxy spectrum is dominated by these strong emission lines. The continuum is marginally detected in the blue channel of the last KCWI observations ($\lesssim 5000$~\AAA) but only emission lines are detected in the red channel. This is expected given the $\gtrsim20$ to 1 SN-galaxy flux ratio at redder wavelengths in the latest KCWI observation. Importantly, \NIIlong is only marginally ($\sim 3\sigma$) detected. 

The \OIII$\lambda4363$ feature is not detected so we must rely on empirical strong-line metallicity estimates. We adopt 
$\log_{10}(\rm O/\rm H)+12 \lesssim 8$~dex based on $\log_{10}$([\ion{N}{2}]/H$\alpha$) $< -1.8$~dex. This corresponds to $\log_{10}(Z_{\rm host} / Z_\odot) \lesssim -0.9$~dex using the Solar abundances from \citet{asplund2021}. The low host-galaxy metallicity is supported by where it lies in the relations between metallicity and galaxy luminosity \citep[e.g., ][]{arcavi2010, sanders2013}. The faintness of the host ($M_{g,r} \approx -13$~mag) places it outside the bounds of the calibration samples for these relations, but extrapolating to such faint dwarfs (as done by \citealp{anderson2018} for the host of the metal-poor SN~II~2015bs) produces a similar global metallicity estimate of $\log_{10} (Z_{\rm host} / Z_\odot) = -1.3\pm0.3$~dex. 

We derive integrated host-galaxy properties by fitting the host-galaxy photometry from the Galaxy Evolution Explorer \citep[GALEX; ][]{martin2005, morrissey2007}, SDSS \citep{york2000}, the Wide-field Infrared Survey Explorer \citep[WISE; ][]{wright2010} using \textsc{Fast} \citep{fast} with free parameters for the total stellar mass, stellar age, internal extinction, star-formation rate (SFR), and the star-formation history modeled as a delayed exponential. Parameters with meaningful constraints are included in Table~\ref{tab:hosttbl}. The low mass of the galaxy will be discussed further in \S\ref{subsec:discuss.metals}.

\begin{deluxetable*}{lcccccc}
    
    \tablecaption{Median $3\sigma$ limits on pre-explosion outbursts and variability from survey photometry. The ATLAS and ZTF light curves span $\approx 8$ and $\approx 5$ years prior to explosion, respectively. The light curve is shown in Appendix~\ref{app:extra}.
    \label{tab:arxiv}
    }

    \tablehead{
    \colhead{Survey/Filter} & \multicolumn{3}{c}{Single-Epoch} & \multicolumn{3}{c}{30-day Average} \\
     & \colhead{\fnu [\uJy]} & \colhead{App. mag.} & \colhead{Abs. mag.} & \colhead{\fnu [\uJy]} & \colhead{App. mag.} & \colhead{Abs. mag.}
    }

    \tablecolumns{5}

    \startdata
    ZTF-g & 38.2 & 19.9 & -12.4 & 4.7 & 22.2 & -11.8 \\
    ZTF-r & 38.5 & 19.9 & -12.4 & 4.8 & 22.2 & -11.8 \\
    ZTF-i & 44.8 & 19.8 & -14.2 & 7.7 & 21.7 & -12.3 \\
    ATLAS-c & 36.0 & 20.0 & -14.0 & 7.6 & 21.7 & -12.3 \\
    ATLAS-o & 38.5 & 19.9 & -14.1 & 8.2 & 21.6 & -12.4 \\ 
%    ASASSN-g \\
    \enddata
\end{deluxetable*}

\vspace{-1cm}
\section{Photometric Properties}\label{sec:phot}

\begin{figure*}
    \centering
    \includegraphics[width=\linewidth]{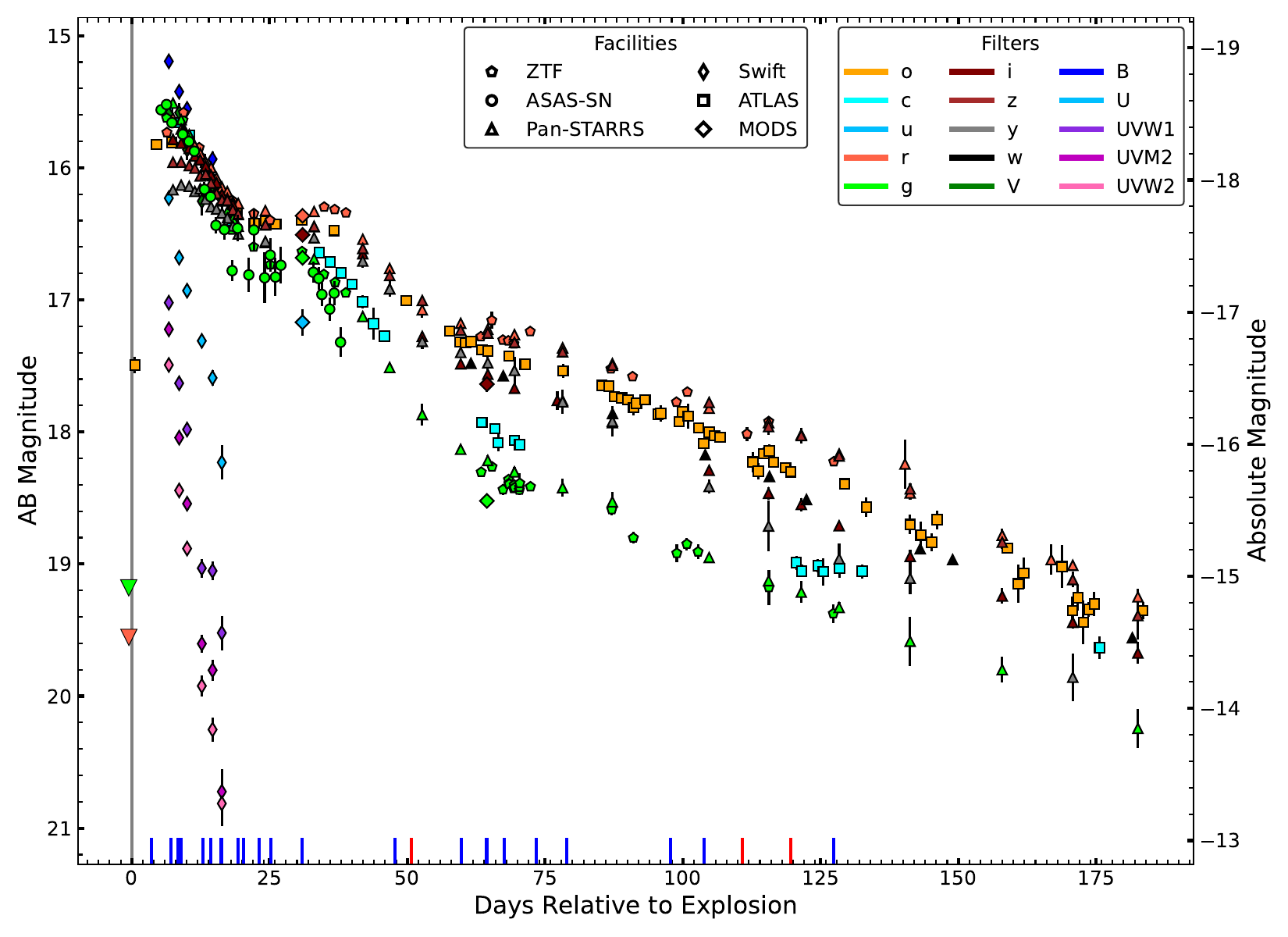}
    \caption{Extinction-corrected light curves of \name. The same colors and symbols are used throughout the manuscript. Inverted triangles show $5\sigma$ non-detections $\sim 0.5$~days before explosion from ZTF. Blue and red ticks along the lower axis denote optical and NIR spectroscopic observations, respectively.}
    \label{fig:allphot}
\end{figure*}

\subsection{Pre-Explosion and Discovery}\label{subsec:phot.disc}

\name was discovered by ATLAS on MJD~60223.58. The public ZTF data stream reports a stringent non-detection $\approx 1$~day prior so we adopt the midpoint of these two observations, MJD~$60223.0\pm0.5$, as the estimated explosion epoch, where the uncertainty spans the full range of possible explosion times.
\name did not experience any detectable outbursts in the $\approx 8$~years prior to explosion. Assuming an outburst would produce enhanced flux for more than a single night, we also compute 30-day averages of the single-epoch fluxes to better constrain fainter events. The 30-day window is motivated by objects with observed pre-explosion variability \citep[e.g., ][]{jg2022, hiramatsu2023} but modifying the length of the averaging window does not affect our conclusions. The median limits on pre-explosion luminosity are given in Table~\ref{tab:arxiv}.

The time required for the light curve to reach peak brightness is correlated with the progenitor radius \citep[e.g., ][]{rabinak2011, morozova2016}. The rising light curve is poorly sampled with only a single ATLAS $o$-band measurement but we synthesize $g$- and $o$-band fluxes from the first SNIFS observations $\approx 2.5$~days after discovery ($\approx 3.5$~days after explosion). We measure a $g$-band rise time of $5\pm1$~days which is only slightly shorter than typical \sneii \citep[e.g., ][]{gonzalez2015}. Using the relation between $g$-band rise time and progenitor radius derived by \citet{morozova2016} we find a pre-explosion radius of $\Rstar \approx 350~\Rsun$, albeit with large uncertainties and the caveat that the relation was calibrated to SNe~IIP and may not accurately describe fast-evolving \sneii.

\subsection{Rapid Evolution}\label{subsec:phot.early}

\begin{figure}
    \centering
    \includegraphics[width=\linewidth]{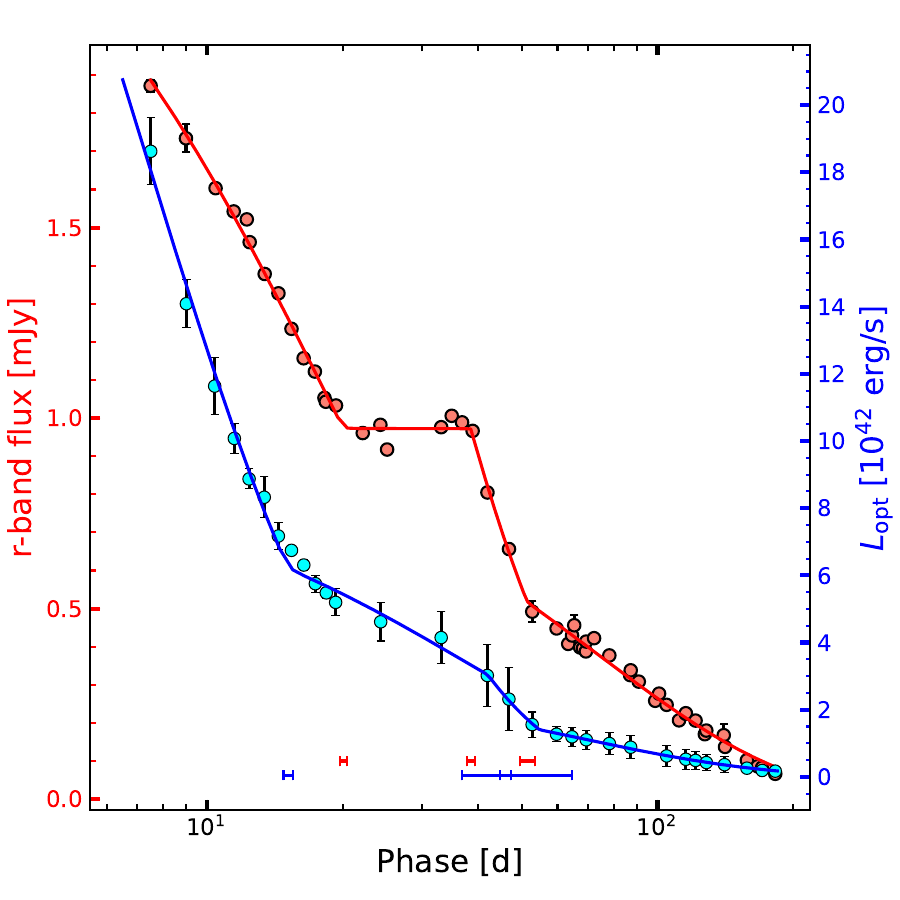}
    \caption{Piece-wise fits to the $r$-band fluxes (red, left scale) and the derived $L_{\rm opt}$ (blue, right scale). Colored error bars along the bottom show the estimated transition times between the different phases of light curve evolution (see \S\ref{subsec:phot.early}).}
    \label{fig:plateau}
\end{figure}

\name peaked at $B/g \approx 15.5$~mag corresponding to $M_{g/B} \approx -18.5$~mag (Fig.~\ref{fig:allphot}), more luminous than most \sneii \citep[e.g., ][]{anderson2014, valenti2016}, but consistent with metal-poor dwarf galaxies hosting overluminous \sneii \citep[e.g., ][]{gutierrez2018, scott2019}. The most notable aspect of the light curve is the short plateau duration seen in Fig.~\ref{fig:allphot} spanning $\sim 10$~days and $\sim 15$~days in the $g$- and $r$-bands, respectively. Fig.~\ref{fig:plateau} shows a piece-wise fit \citep[e.g., ][]{sanders2015} to the $r$-band Pan-STARRS and ZTF observations. The fit divides the evolution into 3 phases: the shock-cooling phase ($\Delta t_{\rm sc}$), the plateau phase ($\Delta t_{\rm pl}$), and the transition from the plateau to the radioactive tail ($\Delta t_{\rm tr}$). There are 4 corresponding exponential decay coefficients: $\beta_{\rm sc}$ during the shock-cooling phase, $\beta_{\rm pl}$ during the plateau, $\beta_{\rm tr}$ during the transition, and $\beta_{\rm tail}$ for the radioactive tail. The values are given in Table~\ref{tab:photfit}.\footnote{The fitted decay coefficients can be converted to the light curve decay parameters introduced by \citet{anderson2014}: $s_1\equiv \Delta m_{\rm sc}$, $s_2\equiv \Delta m_{\rm pl}$, and $s_3\equiv \Delta m_{\rm tail}$.} 

We also compute the optical pseudo-bolometric fluxes using the $grizy$ Pan-STARRS photometry assuming the emission can be approximated by a blackbody. The combined optical+NIR spectra in \S\ref{sec:spex} confirm this is an acceptable assumption even at $\sim 100$~d after explosion. We exclude $r$-band observations $\geq 50$~days after explosion to prevent the strong \Ha emission from biasing the measured luminosities. The $L_{\rm opt}$ light curve is shown alongside the $r$-band light curve in Fig.~\ref{fig:plateau} and piece-wise fit parameters are included in Table~\ref{tab:photfit}. The short ($\approx 20$~d) plateau duration is not due to the strong \Ha emission in the $r$-band. 

The duration of the plateau primarily depends on H envelope mass (e.g., \citealp{kasen2009}). Increased explosion energy and decreased \Mni can extend or shorten the plateau by $\lesssim 10\%$. Applying the scaling relations of \citet{goldberg2019} and \citet{fang2024b} to the $r$-band plateau of \name, we estimate $\Menv = 0.5-1.5~\Msun$. This \Menv estimate is consistent with the modeling estimates of $\Menv \sim 1.7~\Msun$ by \citet{hiramatsu2021} for their sample of 3 short-plateau \sneii (SNe 2006Y, 2006ai, and 2016egz). However, \citet{martinez2022a, martinez2022b, martinez2022c} infer a much higher $M_{\rm env} \sim 7~\Msun$ for SN~2006ai despite measuring a similarly short plateau duration ($\approx 40$~days) as \citet{hiramatsu2021}. These discrepancies highlight the intrinsic difficulty of inferring physical parameters from observables, especially for these faster-evolving \sneii.

\begin{figure}
    \centering
    \includegraphics[width=\linewidth]{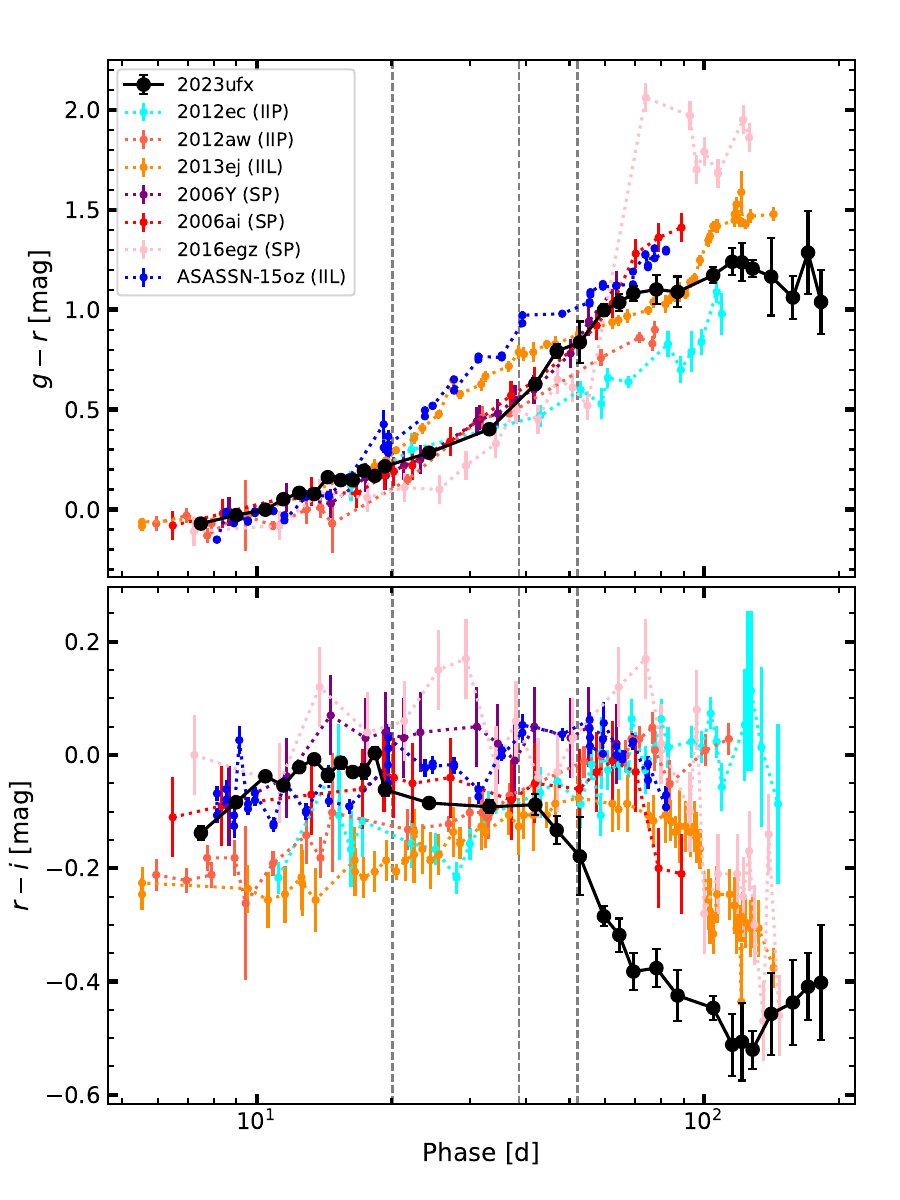}
    \caption{Extinction-corrected color evolution of \name compared to other \sneii. Gray vertical lines mark the different stages of evolution (see \S\ref{subsec:phot.early}).}
    \label{fig:colors}
\end{figure}

Fig.~\ref{fig:colors} shows that \name initially had a similar color evolution to other \sneii, including short-plateau, IIL, and IIP SNe. The color similarity ends when \name enters its early radioactive-decay/nebular phase and develops a strong \Ha emission line, leading to the steadily bluer $r-i$ colors at $\gtrsim 60$~d after explosion. The strong \Ha emission makes the $g-r$ color very red, but the ``intrinsic'' color without \Ha is relatively bluer at $\gtrsim60$~d than most \sne. This is probably due to the early nebular transition revealing a hotter core. The early \emph{Swift} colors (not shown) are indistinguishable from other \sneii, consistent with shock-cooling producing the early light curve. Overall, the lack of color deviations from normal \sneii agrees with the models of \citet{dessart2013}, where metallicity effects are balanced by trade-offs between a smaller progenitor size (faster expansion cooling) and decreased metal-line blanketing at $\lesssim 5000$~\AAA (bluer colors).

\begin{table*}[]
    \centering
    \begin{tabular}{lrrrrrrr}\hline\hline
             & $\Delta t_{\rm sc}$ & $\Delta t_{\rm pl}$ & $\Delta t_{\rm tr}$ & $\log_{10}\beta_{\rm sc}$ & $\log_{10}\beta_{\rm pl}$ & $\log_{10}\beta_{\rm tr}$ & $\log_{10}\beta_{\rm tail}$ \\
             & & & & $\Delta m_{\rm sc}$ & $\Delta m_{\rm pl}$ & $\Delta m_{\rm tr}$ & $\Delta m_{\rm tail}$ \\
           % & [d] & [d] & [d] & & [mag/d$^{-1}] & \\
    \hline
    $g$-band  & $21.7 \pm 1.1$ & $13.8\pm2.3$ & $22.2\pm6.0$ & $-1.19\pm0.02$ & $-1.99\pm0.27$ & $-1.23\pm0.08$ & $-1.80\pm0.11$ \\
    & & & & $7.1\pm0.3$ & $1.1\pm0.7$ & $6.4\pm1.2$ & $1.7\pm0.4$ \\
     \hline
    $r$-band & $20.1\pm0.4$ & $18.4\pm0.7$ & $12.9\pm1.8$ & $-1.28\pm0.01$ & $-4.4\pm4.9$ & $-1.31\pm0.05$ & $-1.86\pm0.02$ \\
            &  &  &  &  $5.7\pm0.2$ & $0.0\pm0.1$ & $5.3\pm0.5$ & $1.5\pm0.1$ \\
    \hline
    $o$-band & $21.6\pm0.8$ & $13.5\pm1.6$ & $19.4\pm2.9$ & $-1.46\pm0.02$ & $-9.2\pm3.3$ & $-1.43\pm0.04$ & $-1.82\pm0.02$ \\
     & & & & $3.8\pm0.2$ & $0.0\pm0.1$ & $4.1\pm0.4$ & $1.6\pm0.1$ \\
     \hline
    $L_{\rm opt}$ & $15.1\pm0.4$ & $26.9\pm5.2$ & $12.7\pm8.4$ & $-0.85\pm0.02$ & $-1.58\pm0.08$ & $-1.22\pm0.15$ & $-1.80\pm0.12$ \\
     & & & & $15.2\pm0.5$ & $2.9\pm0.5$ & $6.6\pm2.3$ & $1.7\pm0.5$\\

    \hline\hline
    \end{tabular}
\caption{Parameters from fitting a piece-wise decay to the optical fluxes and blackbody luminosities (Fig.~\ref{fig:plateau}). Durations ($\Delta t_{\rm sc}$, $\Delta t_{\rm pl}$, $\Delta t_{\rm tr}$) are given in rest-frame days. The second row converts the $\log_{10}\beta$ exponential decay coefficients into the corresponding magnitudes per 100 days ($\Delta m$).}
    \label{tab:photfit}
\end{table*}

\subsection{$^{56}$Ni Mass}\label{subsec:phot.late}

The post-plateau light curve evolution traces energy input from the decay of ${^{56}\rm{Co}\rightarrow^{56}\rm{Fe}}$. We use Eq. 2 from \citet{hamuy2003} to estimate the synthesized Ni mass and a Monte Carlo method to incorporate the photometric and distance uncertainties. We find $M_{\rm Ni56} = 0.13\pm0.04$~\Msun, although the $L_{\rm opt}$ decay ($\Delta m_{\rm tail} \sim 1.4$~mag/100d) is steeper than expected for pure $^{56}$Co decay powering the observed emission. This $^{56}$Ni mass is above-average for \sneii but within the plausible range \citep[e.g., ][]{hamuy2003, pejcha2015, anderson2019} and roughly agrees with the correlation between plateau luminosity and $^{56}$Ni mass from \citet{muller2017}. The measured \Mni is more consistent with the SNe~Ib/IIb/Ic sample of \citet{meza2020}, albeit still above their average values.

\vspace{0.2cm}
\section{Spectroscopic Evolution}\label{sec:spex}

\begin{figure*}
    \centering
    \includegraphics[width=1\linewidth]{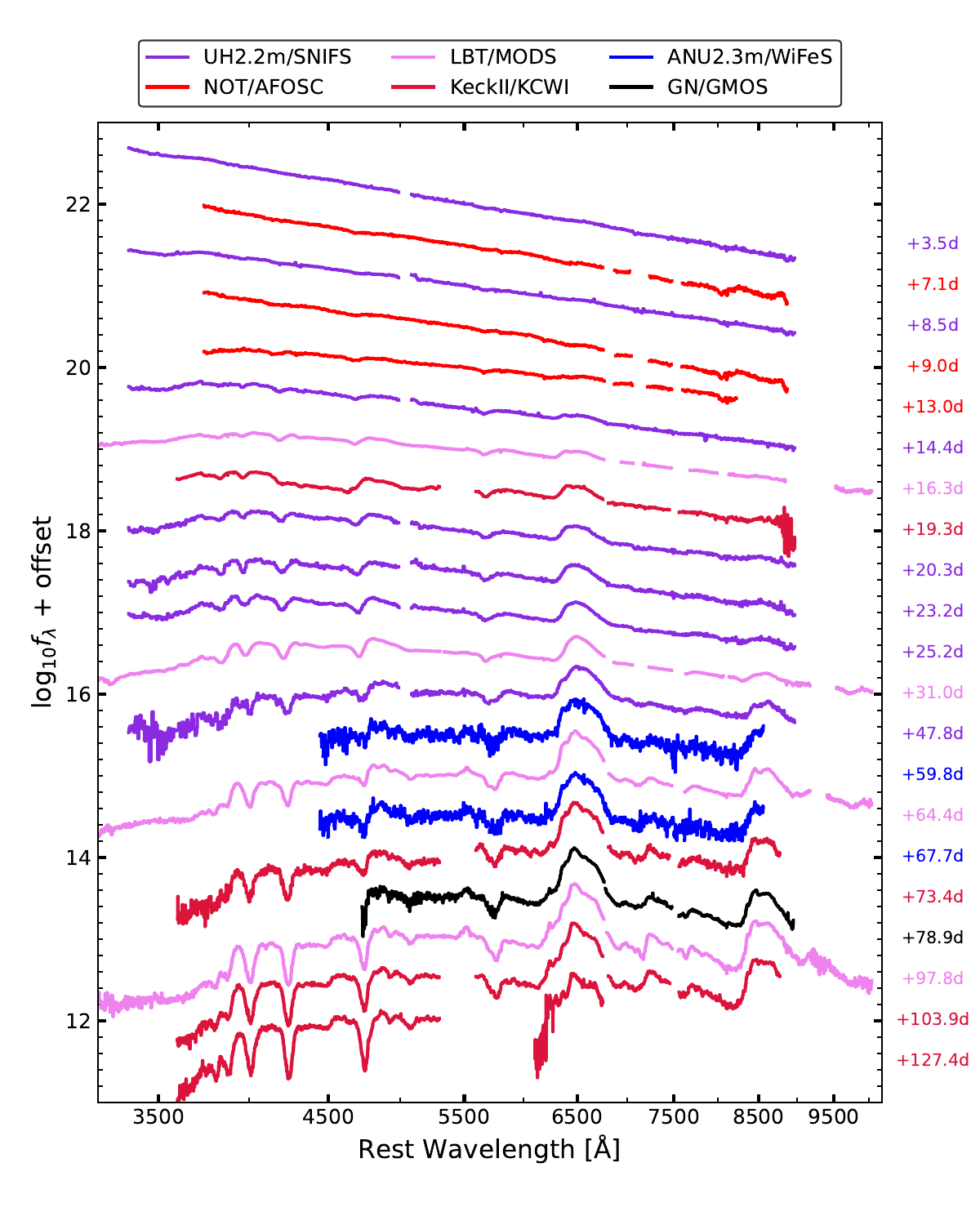}
    \caption{Optical spectra of \name, color-coded by instrument and smoothed for visual clarity. Regions with significant telluric or instrumental artifacts are masked. Phases along the right axis are given relative to \texp (\S\ref{subsec:phot.disc}).
    }
    \label{fig:allspex}
\end{figure*}

\begin{figure}
    \centering
    \includegraphics[width=\linewidth]{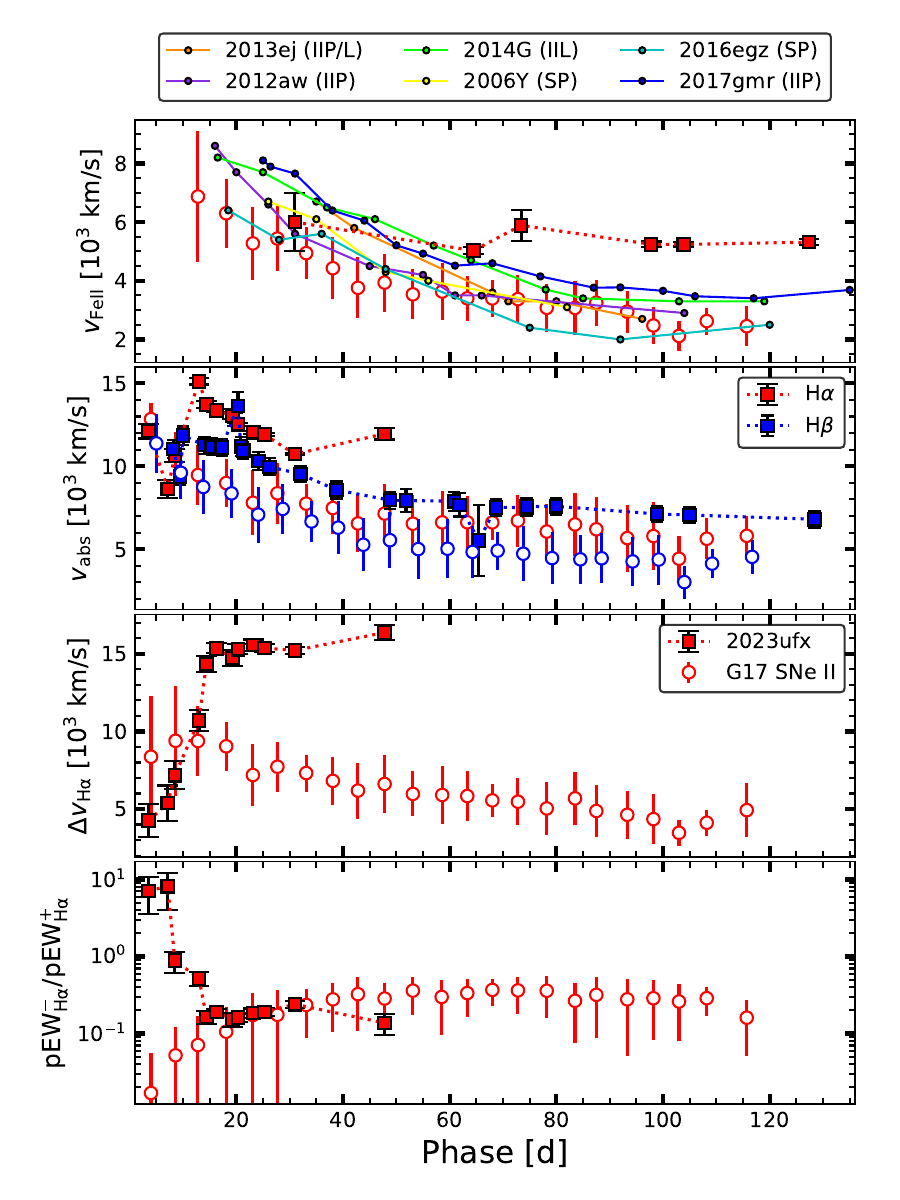}
    \caption{ \ion{Fe}{2}5169\AAA absorption velocity (top), \Ha and \Hb absorption velocities (upper middle), \Ha emission FWHM (lower middle), and the pEW ratio of \Ha absorption over emission (bottom). The ensemble \snii results of \citet[][open circles]{gutierrez2017a} are shown in all panels for comparison and the \FeIIa velocities for some well-studied \sneii are included in the top panel. References are provided in Table~\ref{tab:reftable}.}
    \label{fig:Hevol}
\end{figure}

\subsection{Early Spectra}\label{subsec:spex.early}

The spectroscopic time-series shown in Fig.~\ref{fig:allspex} spans $\sim3-130$~days after explosion. The spectra obtained $\lesssim 15$~days after explosion are almost perfect blackbody curves with only minor deviations from weak, broad spectral features, confirming that the early light curve traces the shock-cooling emission \citep[e.g., ][]{morag2024}. Some \sneii show narrow emission lines in the early ($\lesssim 10$~d) spectra \citep[e.g., ][]{bruch2023, jg2023} due to the shock breakout radiation pulse photo-ionizing the CSM near the stellar surface \citep[e.g., ][]{khazov2016, kochanek2019}. No such features are seen in the early spectra of \name. This can be translated into a constraint on the mass-loss rate at death of $\Mdot \lesssim 10^{-3}$~\Msun/yr \citep[e.g., ][]{dessart2017, boian2019, jg2024} due to the lack of any narrow H, \ion{He}{1}, and \ion{He}{2} in the spectra.

\subsection{Photospheric Spectra}\label{subsec:spex.phot}

\begin{figure*}
    \centering
    \includegraphics[width=\linewidth]{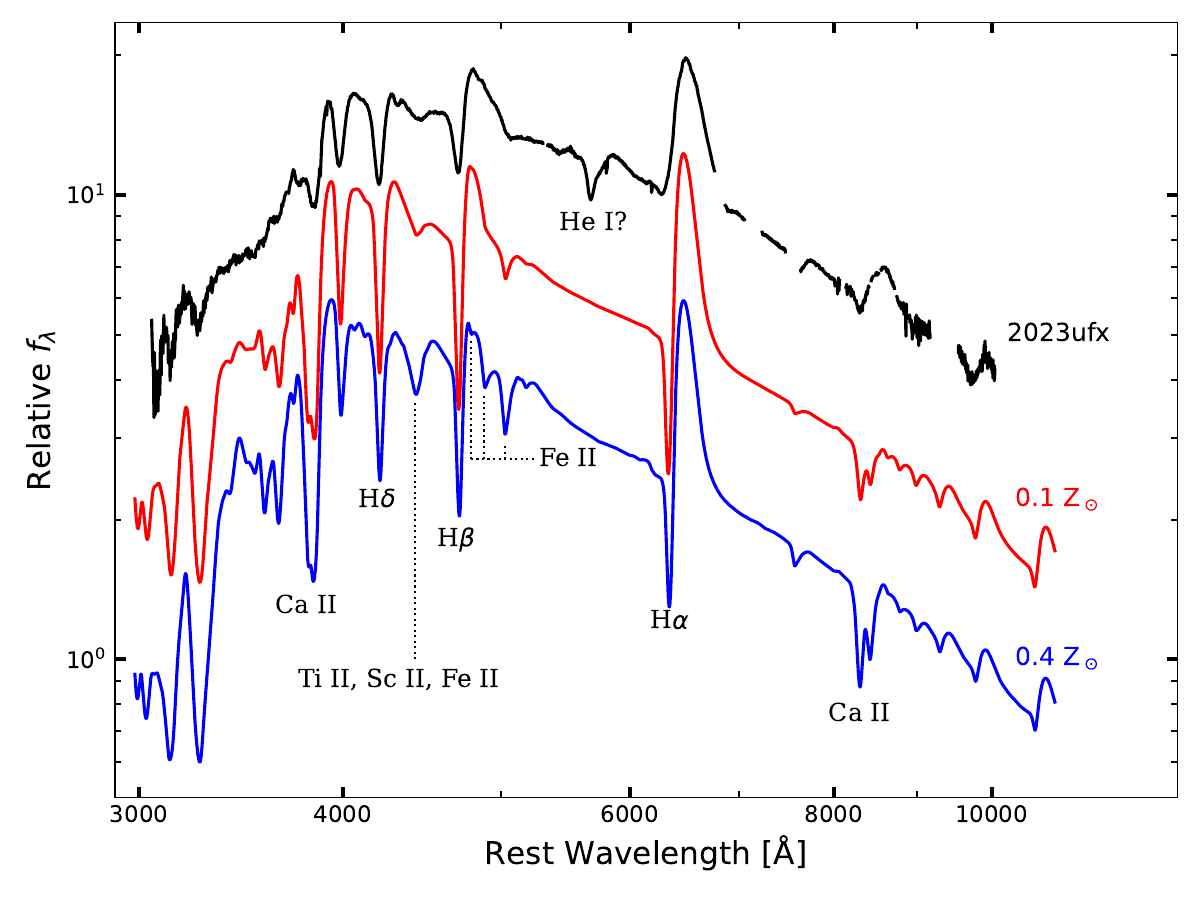}
    \caption{Comparison of the +31~d MODS spectrum with the $0.4~Z_\odot$ (blue) and $0.1~Z_\odot$ (red) model spectra from \citet{dessart2013} at the same phase. Spectra are aligned to the \Hb blue shift of \name to ease visual comparisons. Important spectral features are marked.}
    \label{fig:photspex}
\end{figure*}

\begin{figure}
    \centering
    \includegraphics[width=\linewidth]{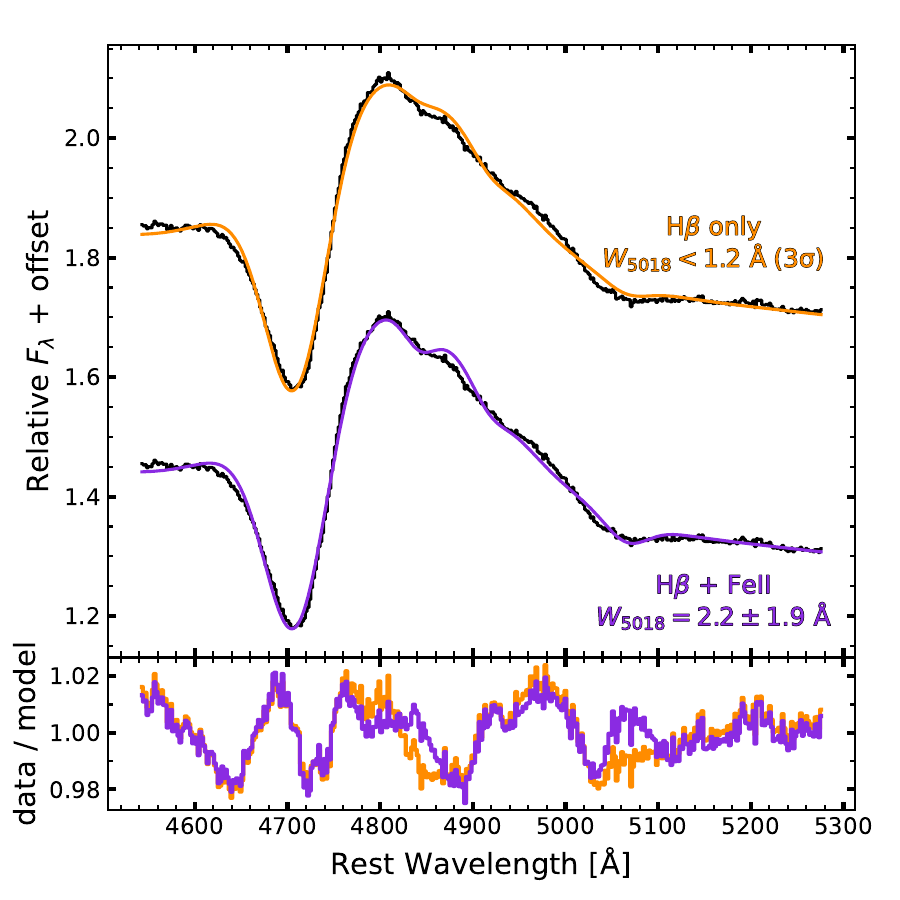}
    \caption{Comparison of two methods for estimating the \ion{Fe}{2}$\lambda5018$ pEW in the high-quality +31~d MODS spectrum (Fig.~\ref{fig:photspex}) discussed in \S\ref{subsec:spex.phot}. The residuals are shown in the lower panel following the same color scheme.}
    \label{fig:FeIIlimit}
\end{figure}

Days $\approx 20-40$ are characterized by spectra typical of \sneii during the plateau phase, including a broad P-Cygni \Ha profile and absorption in the higher-order Balmer transitions. The photospheric spectra show minimal \Ha absorption and above-average velocities, in agreement with other fast-evolving (short-plateau and IIL) SNe~II \citep[e.g., ][]{gutierrez2014, hiramatsu2021}. Fig.~\ref{fig:Hevol} shows the evolution of the \Ha and \Hb features compared to the \sneii sample of \citet{gutierrez2017a}. Overall, the \Ha and \Hb velocities of \name are slightly faster than normal \sneii during the plateau phase.

The starkest discrepancy with other \sneii is the nearly complete absence of any metallic absorption features. Fig.~\ref{fig:photspex} compares the high-quality MODS spectrum obtained +31~d after explosion to the $Z$-dependent model spectra of \citet{dessart2013}. The \ion{Fe}{2} and \ion{Ca}{2} features in \name are weaker than even the $0.1~Z_\odot$ model spectrum at a similar phase.

We attempt two methods for deriving an upper limit on the pseudo-equivalent width (pEW) $W_\lambda$ of \ion{Fe}{2}$\lambda5018$ in the +31~d MODS spectrum. The first method is purely empirical and based upon Eq.~4 from \citet{leonard2001}, 

\begin{equation}\label{eq:ewlim}
    W_{\lambda}(N\sigma) = N\times \Delta I\sqrt{W_{\rm line} \Delta X},
\end{equation}
\noindent for a $N\sigma$ limit where $\Delta I$ is the root-mean-square (RMS) of the normalized continuum, $W_{\rm line}$ is the width of the spectral feature, and $\Delta X$ is the spectral bin size. For the MODS spectrum, $\Delta X = 2.5~\AAA$ and we adopt $W_{\rm line} = 64~\AAA$ corresponding to a \ion{Fe}{2} velocity FWHM of $3000~\kms$, slightly higher than the $\approx 2500~\kms$ FWHM measured for \ion{Fe}{2}$\lambda5018$ at $>60$~d after explosion.

\ion{Fe}{2}$\lambda5018$ sits atop the \Hb P-Cygni profile so we must first remove its contribution to estimate $\Delta I$. We fit a simple double-Gaussian model to the profile with one absorption component and one emission component which decently replicates the observed profile, as shown in Fig.~\ref{fig:FeIIlimit}. Normalizing by this continuum we find a $N\sigma$ upper limit on the \ion{Fe}{2}$\lambda5018$ pEW of $W_{5018}(N\sigma) < N\times 0.4~\AAA$.

\begin{figure}
    \centering
    \includegraphics[width=\linewidth]{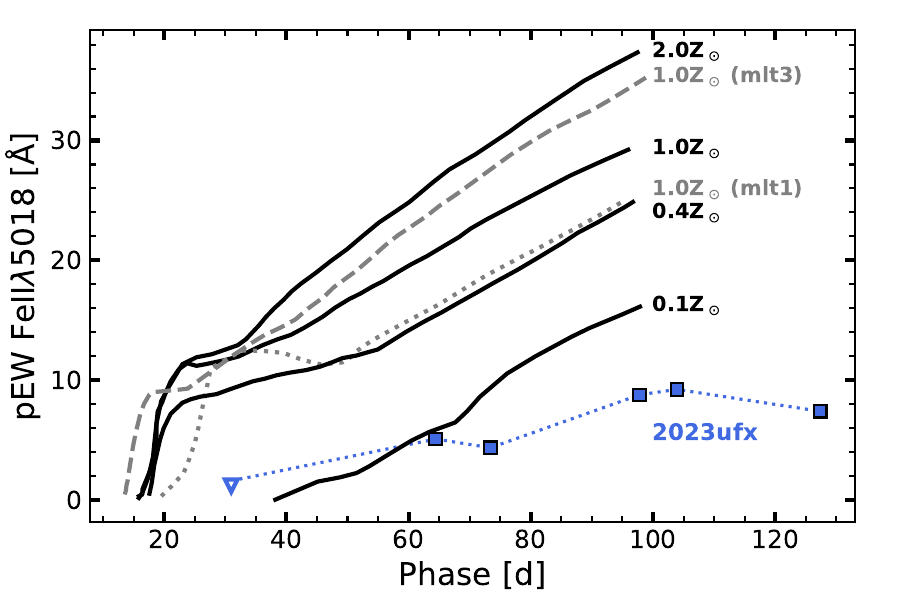}
    \caption{The evolution of the \ion{Fe}{2}$\lambda5018$ pEW in \name (blue) compared to the models from \citet{dessart2013} at different metallicities (black). Two Solar-metallicity models with different mixing-length prescriptions during stellar evolution, also from \citet{dessart2013}, are shown in gray. The inverted triangle shows the pEW upper limit derived in \S\ref{subsec:spex.phot} and shown in Fig.~\ref{fig:FeIIlimit}.}
    \label{fig:feII5018}
\end{figure}

We also tried to fully model the \Hb P-Cygni profile and \ion{Fe}{2}$\lambda4924,5018,5169$ features using an MCMC routine. This only converges if we fix the velocities of the \ion{Fe}{2} features, otherwise the minimizer consistently prefers solutions with implausibly broad ($\Delta v > 10^4~\kms$) \ion{Fe}{2} lines. The best-fit model has $W_{5018} = 2.2\pm1.9~\AAA$ and it is included in Fig.~\ref{fig:FeIIlimit}. The uncertainty is large due to the strong covariance between $W_{5018}$ and the strength and width of the \Hb emission. 

We adopt a conservative upper limit of $W_{5018} < 1.6~\AAA$ for the +31~d spectrum, corresponding to an empirical $4\sigma$ limit, which qualitatively agrees with the MCMC results. This value is included in Fig.~\ref{fig:feII5018} and confirms the metal-poor nature of \name. The low signal to noise ratio of the three spectra spanning $37-60$~d precludes meaningful constraints on the \ion{Fe}{2} features. For example, the $3\sigma$ upper limit on the pEW of \ion{Fe}{2}$\lambda5018$ in the SNIFS +48~d spectrum from Eq.~\ref{eq:ewlim} is $\approx 12~\AAA$.

\subsection{Late/Semi-nebular Spectra}\label{subsec:spex.late}

\begin{figure}
    \centering
    \includegraphics[width=\linewidth]{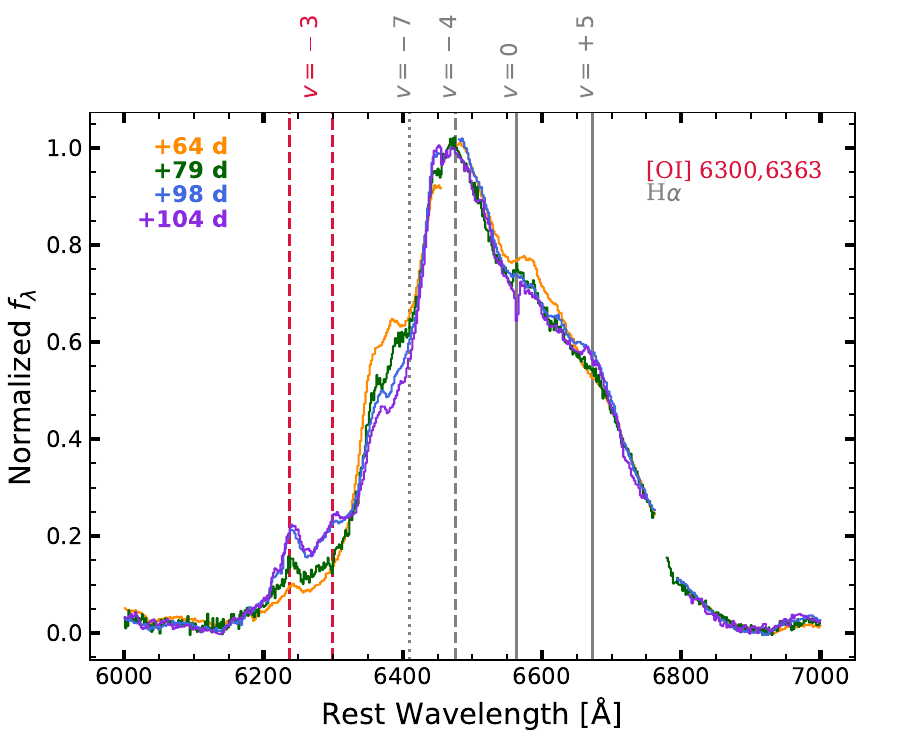}
    \caption{Normalized evolution of the \Ha emission complex spanning $\approx 65-105$~days after explosion. Phases for each spectrum are provided in the upper-left corner. \OI appears at the onset of the radioactive decay phase and strengthens with time. A red-shifted \Ha component appears in the last KCWI high-resolution spectrum -- a comparison to the NIR spectra in Fig.~\ref{fig:paschen} confirms that this feature is not due to low-velocity \ion{He}{1}$\lambda6678$. Labels along the top axis give velocities in $10^3~\kms$ for \Ha (gray) and \OI (red).}
    \label{fig:lateHa}
\end{figure}

\begin{figure}
    \centering
    \includegraphics[width=\linewidth]{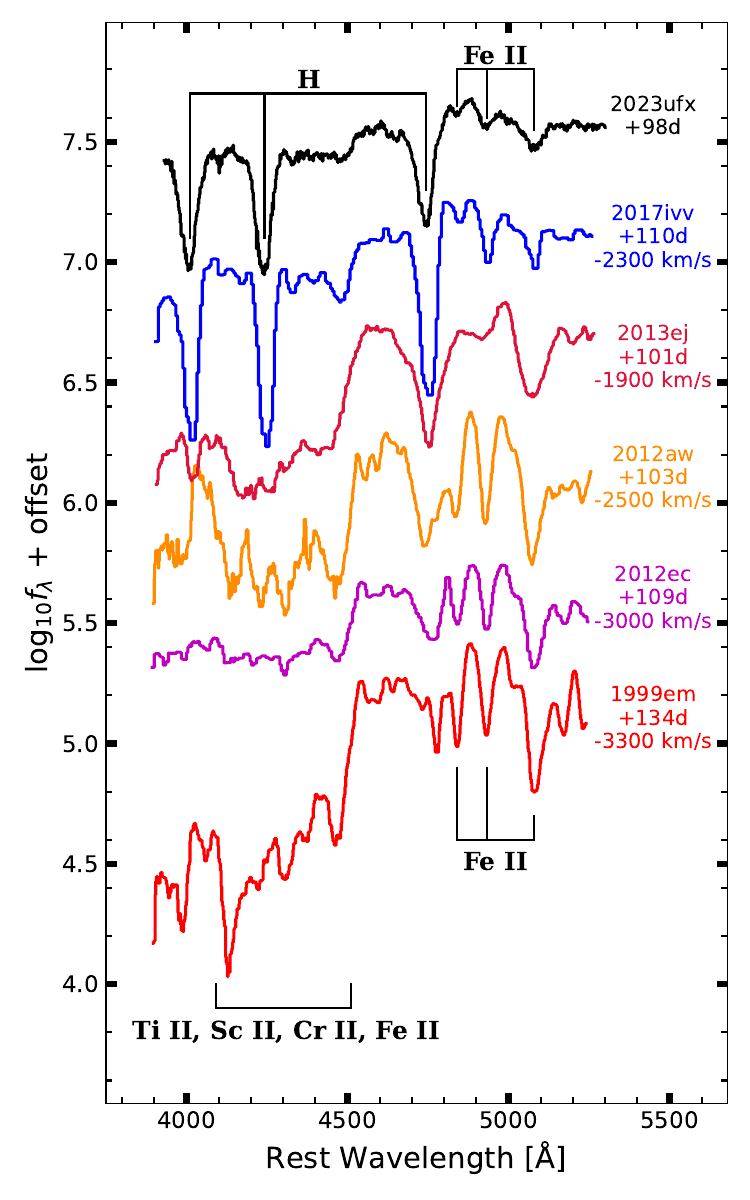}
    \caption{Comparing the high-quality +98~d spectrum to other \sneii in the literature. Each spectrum is aligned to \ion{Fe}{2}$5169$ with the velocity offset given in each label. \name exhibits strong H absorption, weak \ion{Fe}{2} features, and the blend of singly-ionized metals from $\approx 4100-4500$~\AAA is absent. }
    \label{fig:compare_late}
\end{figure}

The spectra $\gtrsim 50$~days after explosion show no evidence for \Ha absorption, consistent with the light curve transitioning into the radioactive-decay phase. The \Ha emission profiles develop an asymmetric, multi-peaked structure with a full-width at half-maximum (FWHM) velocity of $\approx 15\times10^3$~\kms. Fig.~\ref{fig:lateHa} shows the evolution of the \Ha profile. A distinct red component in the \Ha profile appears between the +79 and +98~d spectra which strengthens in the +127~d spectrum. The peak is at $6680$~\AAA, near \ion{He}{1}$\lambda6678$, but the double-peaked Paschen profiles (see below) suggest this is instead red-shifted \Ha moving at $\sim 5000~\kms$. \OI becomes visible and strengthens relative to \Ha with time.

\begin{figure*}
    \centering
    \includegraphics[width=\linewidth]{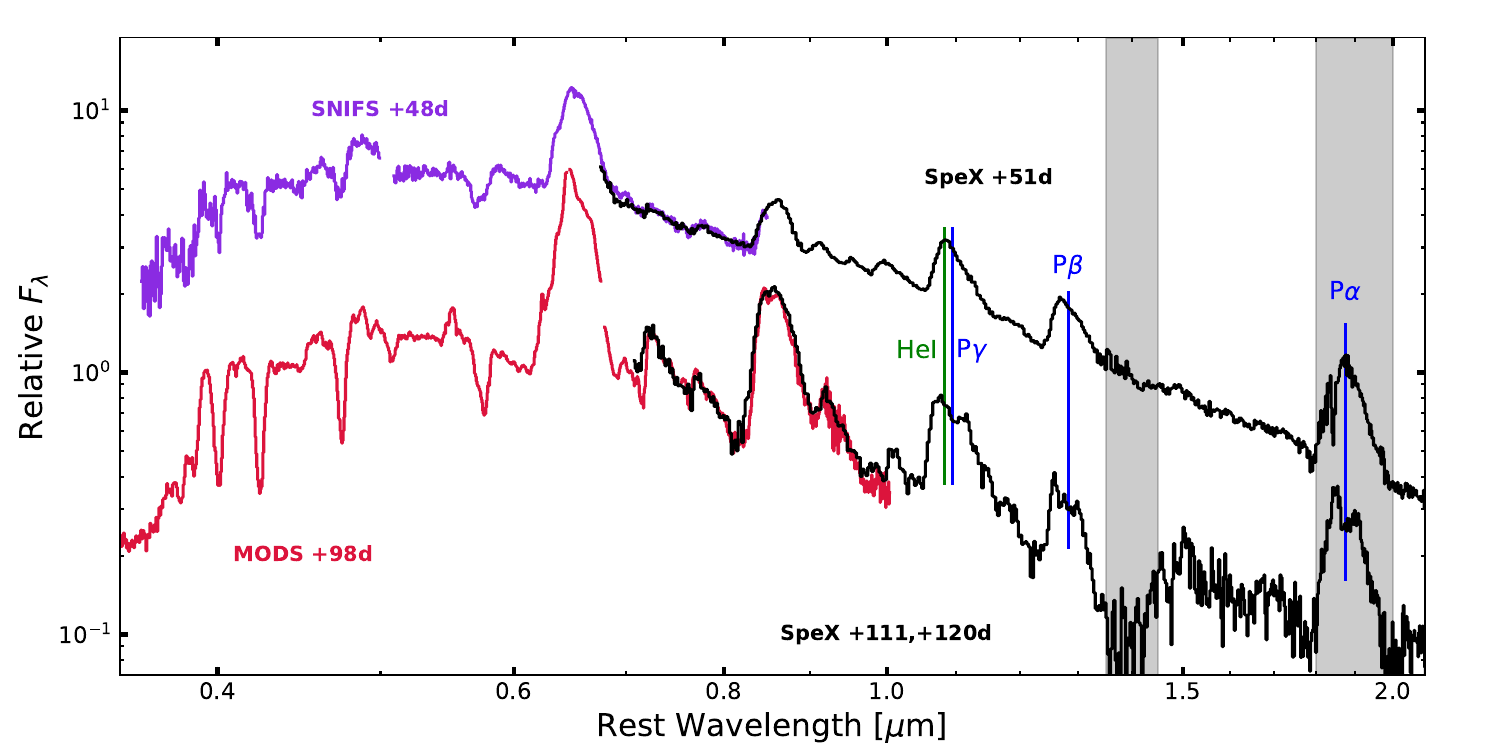}
    \caption{NIR spectra of \name from SpeX (+51~d and $\approx +115$~d, black) shown alongside contemporaneous optical spectra from SNIFS (+48~d, purple) and MODS (+98~d, red). The \Pa, \Pb, \Pg, and \ion{He}{1}~$1.083~\um$ features are marked with vertical colored lines. Regions with significant telluric absorption are shaded gray. }
    \label{fig:irspex}
\end{figure*}

\begin{figure*}
    \centering
    \includegraphics[width=\linewidth]{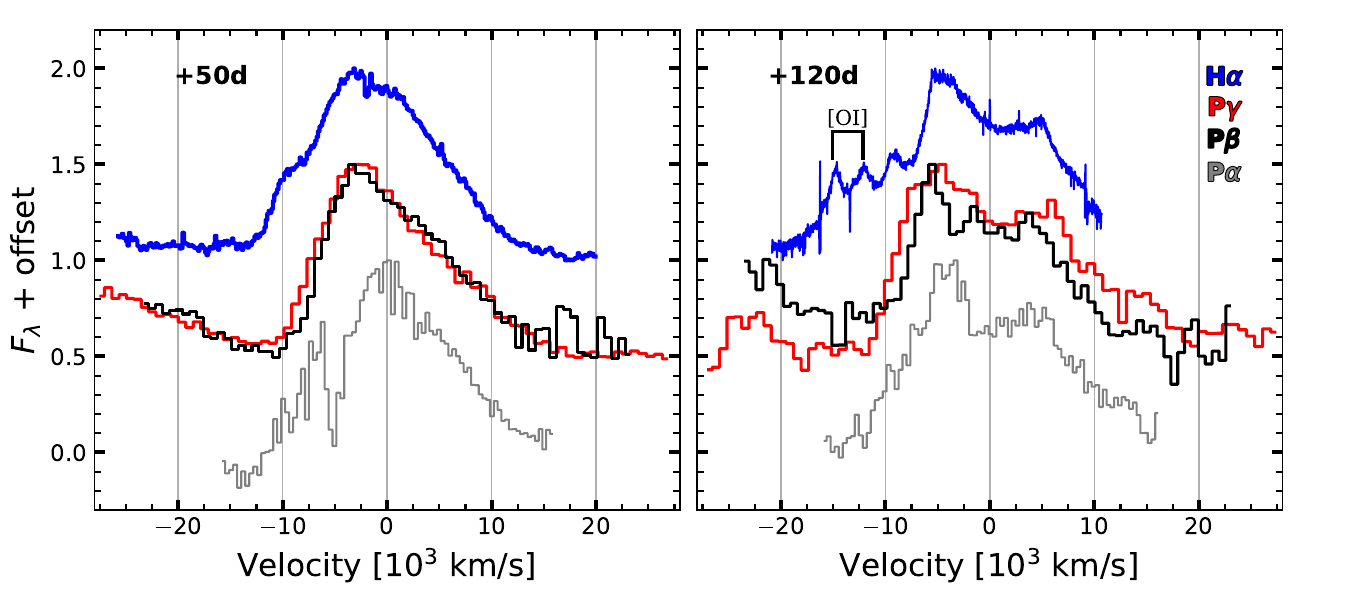}
    \caption{Comparison of the \Ha and Paschen profiles at $\approx +50$~days (left) and $\approx 120$~days (right). While the \Pa line falls in a strong telluric region, its structure is similar to the \Pb and \Pg profiles. The +50~d profiles are single-peaked whereas the +120~d profiles are distinctly double-peaked.}
    \label{fig:paschen}
\end{figure*}

\begin{figure}
    \centering
    \includegraphics[width=\linewidth]{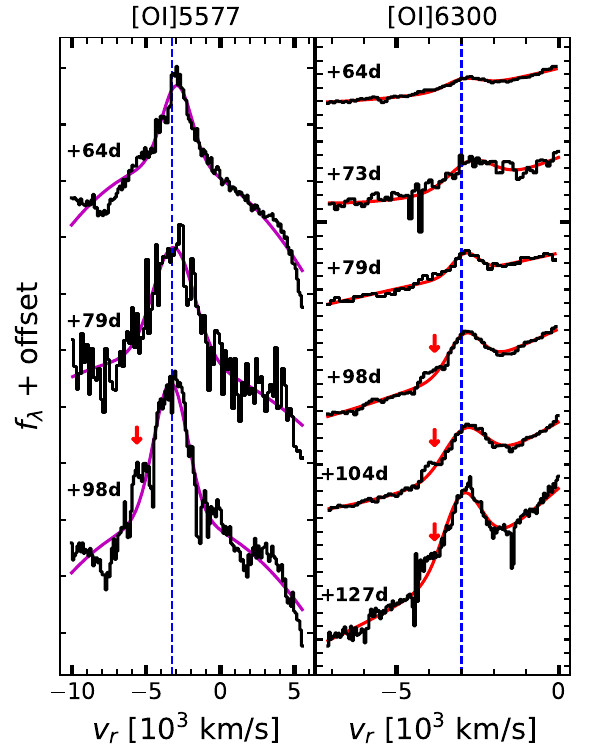}
    \caption{Evolution of the \OI$\lambda5577$ and \OI$\lambda6300$ emission lines at $\approx65-130$~days after explosion. Blue dashed lines show the average velocity shift of each line, $v_{\rm 5577}\approx -3300~\kms$ and $v_{6300}\approx -3000~\kms$. The \OI$\lambda5577$ line is twice as broad ($\approx 2600~\kms$) compared to the \OIa emission ($\approx 1300~\kms$). Red arrows denote potential high-velocity features.}
    \label{fig:Olines}
\end{figure}

\FeII absorption finally appears at $\gtrsim 60$~days and thus must originate in the He layer, confirming that the H envelope was almost entirely devoid of Fe ions. But, as seen in Fig.~\ref{fig:compare_late}, the He-layer \ion{Fe}{2}$\lambda5018$ features are significantly weaker than in other \sneii. In the absence of the metal blends at $\lesssim 4500$~\AAA, \Hg and \Hd remain strong absorption features until the end of our campaign. 

The lack of velocity evolution in \FeIIa agrees with it originating in the He layer but the velocities measured from the absorption-line minima are anomalously high compared to the other \sneii in Fig.~\ref{fig:Hevol}. The high velocities are seen in multiple spectra with an excess of $\approx 2500~\kms$. This will be discussed in \S\ref{subsec:discuss.rot}.

Fig.~\ref{fig:irspex} shows the phase-matched optical and NIR spectra. The near-IR Paschen features exhibit complex velocity structure. We overlay the \Pb and \Pg profiles with \Ha in Fig~\ref{fig:paschen}. Inspecting the compilations of \snii NIR spectra published by \citet{davis2019} and \citet{tinyanont2024}, we find no similarities with the NIR spectra of \name beyond the presence of Paschen emission lines. 

 The shapes of the double-peaked \Pb and \Pg profiles are broadly consistent with the \Ha profile. Fitting the averaged Paschen profile with a double Gaussian model we find velocity shifts $v_1 = -5700\pm900~\kms$ and $v_2 = 3800\pm1200$~\kms. The widths are $\Delta v_1 = 6700\pm800$~\kms and $\Delta v_2 = 11400
 \pm1600~\kms$. The ratio of blue to red flux is $f_{\rm b/r} = 0.63\pm0.13$. If we assume the blue and red components have similar velocities from a common origin (which may be incorrect), the two peaks are separated by $\pm 4800~\kms$ from a bulk velocity of $-1000~\kms$.

 We defer direct comparisons to the nebular spectra models of \citet{jerkstrand2014} until we can obtain spectra deeper into the nebular phase. The early-/semi-nebular spectra do show \OI, which is a diagnostic of the synthesized O mass, but the models are computed for $\gtrsim 200$~days after explosion when the densities are much lower than our observations of \name. Metallicity also likely plays a role in both the envelope and core emission properties given the distinct observational properties of \name outlined so far.  

 The clearest difference between \name and normal \sneii at these phases is the consistent presence of \OIc in Fig.~\ref{fig:Olines}. The ratio of \OI 5577 to 6300 flux ($f_{5577/6300}$) constrains the density and temperature of the emitting material \citep[e.g., ][]{hartigan2004} with typical nebular \sneii exhibiting $f_{5577/6300}\sim 0.01-0.1$ \citep{jerkstrand2014}. We measure the flux ratio in the three spectra that cover both features (GMOS + MODS) finding $f_{5577/6300} = 3.8\pm0.5$, $4.6\pm0.9$, and $2.0\pm0.2$ at +64, +79, and +98~d, respectively. \OIc is seen in late-time spectra of \sneiib \citep[e.,g., ][]{jerkstrand2015} such as SN~2011dh \citep[e.g., ][]{shivvers2013, ergon2014} and $f_{5577/6300} > 1$ implies a hotter and denser He core than normal \sneii, again suggestive of a massive progenitor. The \OIlong flux ratio is below the optically-thin limit ($f_{6363/6300} = 1/3$) at $<100$~d, suggesting photon scattering in the H envelope is suppressing the redder component. The flux ratio rises to $0.8\pm0.1$ by +127~d, approaching the optically-thick limit ($f = 1$) and closer to the expectations for the \OIlong evolution \citep[e.g., ][]{maguire2012}.
 
%\vspace{1.5cm}
\section{Discussion}\label{sec:discuss}

\subsection{A Glimpse at SNe II in the Early Universe}\label{subsec:discuss.metals}

\begin{figure*}
    \centering
    \includegraphics[width=\linewidth]{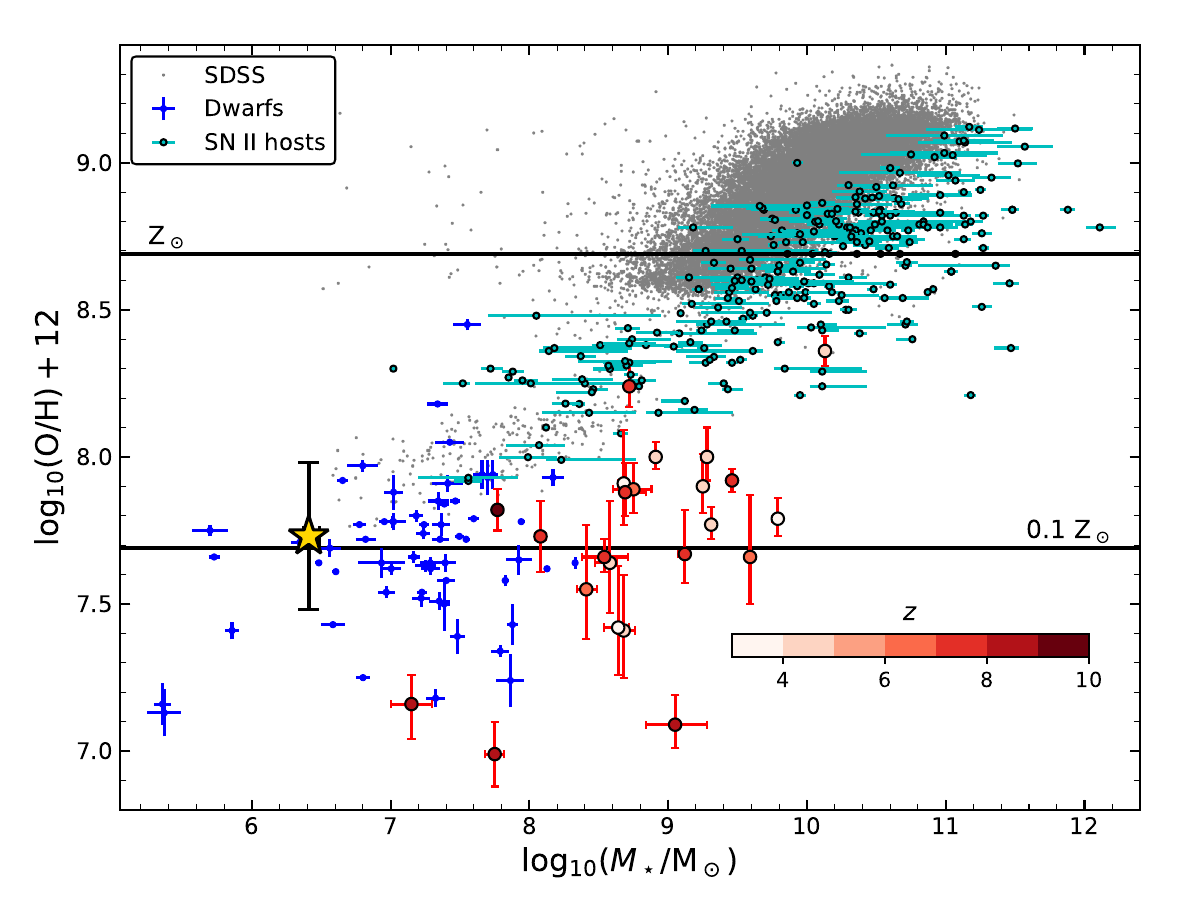}
    \caption{The galaxy mass-metallicity plane discussed in \S\ref{subsec:discuss.metals}. We compare the host-galaxy of \name (gold star) to SDSS (gray points, \citealp{eisenstein2011}), local dwarfs (dark blue points, \citealp{berg2012, hsyu2018}), \snii hosts (light blue points, \citealp{kelly2012, gutierrez2018, taggart2021}), and high-redshift galaxies from \emph{JWST} \citep{nakajima2023, curti2023, heintz2023, morishita2024} with the redshift given by the red colorbar.
    }
    \label{fig:MvsZ}
\end{figure*}

To highlight the uniqueness of \name's environment, we show the galaxy mass-metallicity plane in Fig.~\ref{fig:MvsZ}. We use the luminosity-metallicity relation derived by \citet{berg2012} specifically for low-luminosity galaxies. Using their fit to the `combined+select' sample and $M_B = -13.29\pm0.16$~mag from the SDSS photometry suggests $12+\log_{10}(\rm O/\rm H) = 7.73\pm0.25$~dex ($Z_{\rm host}\sim0.1~Z_\odot$). This value is shown in Fig.~\ref{fig:MvsZ}.

For comparison we include galaxies from SDSS DR8 \citep{eisenstein2011} and \snii hosts from \citet{kelly2012}, \citet{gutierrez2018}, and \citet{taggart2021}. SN metallicities are updated from \citet{graham2019, ganss2022} and \citet{pessi2023a} where possible. We caution that oxygen abundance estimates for \sneii are heterogeneous, with local \sneii benefiting from local/environmental estimates from integral-field observations \citep[e.g., ][]{kuncar2013, kuncar2018, galbany2016}, whereas more distant sources must rely on integrated galaxy properties and luminosity-metallicity relations \citep[e.g., ][]{arcavi2010, sanders2013}. We also include the targeted dwarf galaxies from \citet{berg2012} and \citet{hsyu2018} and a sample of high-$z$ galaxies from \emph{JWST} with `direct' ($T_e$-based) oxygen abundances \citep{nakajima2023, curti2023, heintz2023, morishita2024}.

The combination of low host metallicity and the dearth of metal lines in the photospheric spectra (cf. Figs.~\ref{fig:photspex} and \ref{fig:compare_late}) implies $Z_{\rm SN} < 0.1~Z_\odot$. The iron in the envelopes of RSGs is primordial not nucleosynthetic so this low metallicity directly applies to the progenitor star. Only a handful of extremely metal-poor \sneii have been identified thus far. Besides \name, only SNe~2015bs \citep{anderson2018} and 2017ivv \citep{gutierrez2020} are compelling cases for $Z_{\rm SN} \lesssim 0.1~Z_\odot$ exhibiting both weak \ion{Fe}{2} features and extremely-low host metallicities. The list of metal-weak ($Z \lesssim 0.5~Z_\odot$) \sneii is growing \citep[e.g., ][]{taddia2016, singh2018, boestrom2019, gutierrez2020, zhang2022}, but the local scarcity of such environments inherently limits these opportunities.

\subsection{Mass loss: Stellar winds or binarity?}\label{subsec:discuss.binarity}

The short plateau requires a low \Menv ($\sim 1~\Msun$, \S\ref{subsec:phot.early}) which is difficult to reconcile with typical RSG evolution producing $\gtrsim 5~\Msun$ H envelopes and $\sim 80-150$~day plateaus \citep[e.g., ][]{sukhbold2016, barker2022}. The two mechanisms for removing several \Msun from the envelope prior to collapse are strong, possibly eruptive, stellar winds or stripping by a nearby companion.

Conventional stellar-wind prescriptions \citep[e.g., ][]{dejager1988, vink2001} depend on metallicity as $\Mdot \propto Z^\alpha$ where $\alpha=0.3-0.8$ \citep[e.g., ][]{vink2022} so stellar winds should be $10-50\%$ weaker for the progenitor of \name than for a star at Solar metallicity. Weakened stellar winds at low $Z$ should increase the fraction of massive stars that retain some of their H envelope at death (i.e., the $N_{\rm II}/N_{\rm Ibc}$ ratio; e.g., \citealp{stanway2020}). Thus, a massive star that would have exploded as a SN Ib or IIb in a luminous metal-enriched galaxy may instead produce a short-plateau \snii like \name.

An initially massive progenitor for \name agrees with the above-average \Mni$\sim 0.1~\Msun$ estimated in \S\ref{subsec:phot.late} and the detection of \OIc (\S\ref{subsec:spex.late}). Yet this requires some fine-tuning to produce \name as the star must be massive enough to partially strip the envelope but not so massive as to remove it entirely. Given that \name is just the third \snii with extremely low metallicity, such a fine-tuned scenario seems implausible but is not strictly impossible.

The lack of narrow emission features in the early spectra of \name restrict the mass-loss rate at death to $< 10^{-3}~\Msun$/yr (\S\ref{subsec:spex.early}) which disallows large outbursts or eruptions leading up to collapse. It does not exclude enhanced winds during the main-sequence phase and \S\ref{subsec:discuss.rot} outlines evidence for enhanced rotation in the progenitor of \name. Turbulent motions in the envelope can boost wind mass-loss rates by orders of magnitude \citep[e.g., ][]{kee2021} and rotation-induced turbulence can also dredge up metals from the core into the envelope to increase wind-driven mass loss \citep[e.g., ][]{markova2018}. However, the low Fe abundance in the photospheric phase requires minimal metals in the RSG envelope and excludes strong mixing during the explosion. Theoretical studies are needed to fully capture the connections between turbulence, rotation, and stellar winds at low $Z$.

Binary interaction is the other mechanism for removing several \Msun from the envelope. Stripping from a nearby companion is commonly invoked to explain the sequence of partially-stripped \sneii \citep[e.g., ][]{moriya2016,dessart2024} and can explain the observed occurrence rates \citep[e.g., ][]{eldridge2018}. The primary argument against binarity producing partially or fully stripped CC SNe is the dearth of identified companions. Massive stars have massive companions \citep[e.g., ][]{sana2012} and most will experience mass-transfer or interaction over the course of stellar evolution \citep[e.g., ][]{kobulnicky2007}. The list of SNe II with identified progenitors has steadily grown \citep[e.g., ][]{smartt2004, maund2013, maund2014, fraser2016} yet no binary companions have been identified in a canonical SN II, fast-evolving or otherwise \citep[e.g., ][]{vandyk2023}. Tentative companions have been identified for some SNe~IIb \citep{maund2004,ryder2018,maund2015}, but the Galactic SN IIb Cas A has strict limits on a companion \citep[e.g., ][]{kochanek2018, kerz2019} and most recent Galactic CC SNe did not have a companion at the time of explosion \citep{kochanek2021, kochanek2023}.

This lack of detected companions is difficult to reconcile with the high O/B binarity fraction of $\sim 70\%$ and subsequent 30\% binary fraction of their RSG descendants \citep[e.g., ][]{neugent2020, neugent2021}. The flat mass-ratio and orbital-period distributions of massive-star binaries \citep[e.g., ][]{shenar2022} also disfavors low-mass companions to all \sneii with detected progenitors. If short-plateau SNe II are truly the result of partial stripping from a nearby companion, there are enough SNe~IIP with progenitor detections \citep[e.g., ][]{rodriguez2022} that a distant (i.e., non-interacting) and massive (mass-ratio $q\gtrsim 0.8$) companion could be present in at least one nearby \snii.

\subsection{CSM Interaction}\label{sec:CSM}
While the lack of narrow features in the early spectra (\S\ref{subsec:spex.early}) exclude optically-thick CSM near the progenitor's surface, it does not probe extended optically-thin CSM that contributes a blue continuum to the observed fluxes. This CSM configuration explains the \sneii with above-average luminosities, weak absorption features, and broad emission lines \citep[e.g., ][]{pessi2023a}. Moreover, all \sneii should become interaction-dominated at some point in their evolution \citep[e.g., ][]{dessart2022} which is seen in local \ccsne \citep[e.g., ][]{milisav2012, rizzosmith2023}. 

CSM interaction could explain the above-average peak luminosity and boxy \Ha and \OI profiles seen in the late-time spectra. Yet the CSM interaction models of \citet{dessart2022} predict UV emission commensurate with shock interaction power. The weakest shock power considered by \citet{dessart2022} of $10^{40}~\rm{erg}~\rm s^{-1}$ predicts an early \emph{Swift} $UVW2$ magnitude of $-14$~mag, brighter than the latest \emph{Swift} observations in Fig.~\ref{fig:allphot}. CSM interaction could have started at later times but the interaction models consistently predict a blue continuum which is not seen in the spectra of \name extending to $\lesssim 3500~\AAA$ (cf. Fig.~\ref{fig:photspex} and \ref{fig:irspex}).

Weaker CSM interaction could be present but it appears unlikely to explain the lack of metal features in \name or the double-peaked Paschen lines. The dilution of spectral features by CSM interaction cannot explain the late-time \Hd and \Hg features in Fig.~\ref{fig:compare_late} which are replaced by metal species in other \sne. We will continue monitoring \name as it progresses through the nebular phase to better understand the pre-explosion circumstellar environment.

\subsection{Linking Metallicity, Rotation, and Feedback}\label{subsec:discuss.rot}

The most plausible explanation for the high \Ha, \OI, and \FeIIa velocities is an inherently aspherical explosion \citep[e.g., ][]{wong2013, wong2015, couch2015}, possibly oriented along our line-of-sight. The velocities are an order of magnitude (or more) too high to attribute to the explosion unbinding a binary or neutron star/black hole kicks \citep[e.g., ][]{sweeney2022}. Radiative transfer and photon scattering in partially-nebular ejecta can produce blue-shifted emission features even when the underlying emitting material has zero net velocity \citep[e.g., ][]{jerkstrand2017}. However, this should produce a distinctive evolution back to the rest velocity of the emitting material as the optical depth diminishes, which is not observed.

There is ample evidence that most (or all) core-collapse explosions are inherently aspherical at some level \citep[e.g., ][]{kifonidis2003, mazzali2005, maeda2008, couch2009, lopez2009, chornock2010, wong2013}, but faster rotation may increase asymmetries at low metallicities \citep[e.g., ][]{maeder2000, woosley2006, cano2017}. Stellar winds exert a torque on the star by interacting with the magnetic field (``magnetic braking''; e.g., \citealp{keszthelyi2022}) and reduce the rotation rate. Weaker stellar winds at lower metallicity will decrease angular momentum losses \citep[e.g., ][]{yusof2013, maeder2014}, leading to rapidly-rotating progenitors with more energetic explosions \citep[e.g., ][]{couch2009, papish2011, mosta2015}.

This connection between rotation and metallicity is seen in the environmental properties of the Type Ic SNe originating from massive stars that have lost their H and He layers. Normal SNe~Ic prefer environments with enhanced (super-Solar) metallicity \citep[e.g., ][]{modjaz2011, modjaz2020}, but the energetic broad-lined SNe Ic(-BL) associated with long $\gamma$-ray bursts and relativistic jets \citep[e.g., ][]{izzo2019} occur almost exclusively in metal-poor dwarf galaxies \citep[e.g., ][]{svensson2010, kruhler2015}. From a purely statistical view, normal and broad-lined SNe Ic may originate from completely independent stellar populations \citep[e.g., ][]{fruchter2006, modjaz2020}. 

The one other \snii with a robust metallicity $\lesssim 0.1~Z_\odot$, SN~2015bs \citep{anderson2018}, appears to be a less extreme version of \name. SN~2015bs was more luminous than local \sneii, exhibited fast ejecta velocities, and had a slightly shorter plateau than the median for SNe~II \citep[e.g., ][]{anderson2014, valenti2016}. If the photometric and spectroscopic properties of local SNe~II \citep{anderson2014, valenti2016, gutierrez2017a, gutierrez2017b} are representative of all SNe~II, the binomial probability of the first two metal-poor \sneii consistently exhibiting all three of these properties is $<1\%$.

This also implies metallicity-dependent feedback from CC SNe because of the increased ejecta velocities and prevalence for outflows and jets. Nucleosynthetic yields also depend on progenitor metallicity due to the core electron fraction increasing with metallicity from partially and fully ionized metals. Thus, cosmological simulations must account for changes in CC SN feedback across cosmic time. The same metals ejected by early \sneii will affect the evolution and explosion of subsequent massive stars. 

\section{Summary}\label{sec:summary}

We presented extensive optical and NIR observations of the peculiar \snii 2023ufx. It exploded in the outskirts of a metal-poor ($Z_{\rm host}\sim 0.1~Z_\odot$) dwarf galaxy ($M_\star \sim 10^{6.4}~\Msun$). \name provides a glimpse into the death of massive stars at low $Z$ expected in the early Universe. The plateau phase is short ($\sim 20$~days) indicating a thin ($\sim 1~\Msun$) H envelope. The complete lack of \ion{Fe}{2} in the plateau-phase spectra confirm a progenitor with $Z_{\star} \lesssim 0.1~Z_\odot$. We measure $\Mni \sim 0.1~M_\odot$ from the late-time light curve which suggests a massive progenitor. The late-time H emission lines are bimodal which we attribute to fast outflows ($\sim 5000~\kms$) launched during the explosion, possibly due to a rapidly rotating progenitor. 

The combination of low metallicity and fast evolution make \name a probe of massive-star evolution in the low-$Z$ early Universe. If the nebular spectra confirm a massive progenitor, \name is evidence for a higher fraction of SNe~II at lower metallicities and high redshifts due to weakened stellar winds. Finding fast outflows in one of the two known \sneii at such low metallicities suggests a connection to the rapidly-rotating progenitors of SNe Ic-BL/GRBs.

%% Also note that the akcnowlodgment environment does not support long amounts of text. If you have a lot of people and institutions to acknowledge, do not use this command. Instead, create a new \section{Acknowledgments}.
\vspace{1cm}
\section*{Acknowledgements}

We thank Michelle Tucker, Todd Thompson, and Kris Stanek for useful discussions. We thank Rick Pogge and Paul Martini for discussions about galaxy metallicity estimates and their reliability. 

%%%%%%%%%%%%%%%%
%%% personal and funding ack

Parts of this research were supported by the Australian Research Council Discovery Early Career Researcher Award (DECRA) through project number DE230101069.

C.A. acknowledges support by NASA grants JWST-GO-02114, JWST-GO-02122, JWST-GO-03726, JWST-GO-04436, and JWST-GO-04522. 

This material is based upon work supported by the National Science Foundation Graduate Research Fellowship Program under Grant Nos. 1842402 and 2236415. Any opinions, findings,and conclusions or recommendations expressed in this material are those of the author(s) and do not necessarily reflect the views of the National Science Foundation.

MN is supported by the European Research Council (ERC) under the European Union’s Horizon 2020 research and innovation programme (grant agreement No.~948381) and by UK Space Agency Grant No.~ST/Y000692/1.

M.P. acknowledges support from a UK Research and Innovation Fellowship (MR/T020784/1).

%%%%%%%%%%%%%%%
%%% data and telescope ack

This work is based on observations made with the Large Binocular Telescope. The LBT is an international collaboration among institutions in the United States, Italy, and Germany. LBT Corporation partners are: The University of Arizona on behalf of the Arizona Board of Regents; Istituto Nazionale di Astrofisica, Italy; LBT Beteiligungsgesellschaft, Germany, representing the Max-Planck Society, The Leibniz Institute for Astrophysics Potsdam, and Heidelberg University; The Ohio State University, representing OSU, University of Notre Dame, University of Minnesota and University of Virginia.

This paper used data obtained with the MODS spectrographs built with
funding from NSF grant AST-9987045 and the NSF Telescope System
Instrumentation Program (TSIP), with additional funds from the Ohio
Board of Regents and the Ohio State University Office of Research.

Pan-STARRS is a project of the Institute for Astronomy of the University of Hawaii, and is supported by the NASA SSO Near Earth Observation Program under grants 80NSSC18K0971, NNX14AM74G, NNX12AR65G, NNX13AQ47G, NNX08AR22G, 80NSSC21K1572 and by the State of Hawaii.

The Pan-STARRS1 Surveys (PS1) and the PS1 public science archive have been made possible through contributions by the Institute for Astronomy, the University of Hawaii, the Pan-STARRS Project Office, the Max-Planck Society and its participating institutes, the Max Planck Institute for Astronomy, Heidelberg and the Max Planck Institute for Extraterrestrial Physics, Garching, The Johns Hopkins University, Durham University, the University of Edinburgh, the Queen's University Belfast, the Harvard-Smithsonian Center for Astrophysics, the Las Cumbres Observatory Global Telescope Network Incorporated, the National Central University of Taiwan, the Space Telescope Science Institute, the National Aeronautics and Space Administration under Grant No. NNX08AR22G issued through the Planetary Science Division of the NASA Science Mission Directorate, the National Science Foundation Grant No. AST-1238877, the University of Maryland, Eotvos Lorand University (ELTE), the Los Alamos National Laboratory, and the Gordon and Betty Moore Foundation.

This paper used data obtained with the Infrared Telescope Facility, which is operated by the University of Hawaii under contract 80HQTR19D0030 with the National Aeronautics and Space Administration.

Based on observations made with the Nordic Optical Telescope, owned in collaboration by the University of Turku and Aarhus University, and operated jointly by Aarhus University, the University of Turku and the University of Oslo, representing Denmark, Finland and Norway, the University of Iceland and Stockholm University at the Observatorio del Roque de los Muchachos, La Palma, Spain, of the Instituto de Astrofisica de Canarias. The data presented here were obtained in part with ALFOSC, which is provided by the Instituto de Astrofisica de Andalucia (IAA) under a joint agreement with the University of Copenhagen and NOT.

The automation of the ANU 2.3-metre telescope was made possible through an initial grant provided by the Centre of Gravitational Astrophysics and the Research School of Astronomy and Astrophysics at the Australian National University and partially through a grant provided by the Australian Research Council through LE230100063.


\vspace{5mm}
\facilities{UH:2.2m (SNIFS), NOT (ALOFC), Keck:II (KCWI), IRTF (SpeX), LBT (MODS), ANU 2.3m (WiFeS)}


\software{astropy \citep{astropy:2013, astropy:2018, astropy:2022}, matplotlib \citep{matplotlib}, lmfit \citep{lmfit}, spectres \citep{spectres}, numpy \citep{numpy}, pandas \citep{pandas}, extinction \citep{extinction}, scipy \citep{scipy}, astro-scrappy \citep{astroscrappy}}



\clearpage

\appendix

\section{Data Reduction and Calibration}\label{app:obs}

\subsection{Photometry}
We provide the light curves for \name from the ATLAS ($c$, $o$), ASAS-SN ($g$), ZTF ($g$, $r$), and Pan-STARRS ($g$, $r$, $i$, $z$, $y$) surveys as supplementary material. Additional photometry was obtained with the Neil Gehrels \emph{Swift} observatory using the UltraViolet and Optical Telescope (UVOT) and the Large Binocular Telescope (LBT) Multi-Object Double Spectrograph (MODS, \citealp{pogge2010}). 

For most of the \swift{} epochs, \name was observed with all six UVOT filters \citep{poole08}: $V$ (5425.3 \AA), $B$ (4349.6 \AA), $U$ (3467.1 \AA), $UVW1$ (2580.8 \AA), $UVM2$ (2246.4 \AA), and $UVW2$ (2054.6 \AA). Most UVOT epochs contain at least 2 observations per filter, which we combined into one image for each filter using the HEASoft {\tt uvotimsum} package. We used the {\tt uvotsource} package to extract source counts using a 5\farcs{0} radius region centered on the position of the supernova and background counts using a source-free region with radius of 50\farcs{0}. We then converted the UVOT count rates into fluxes and magnitudes using the calibrations of \citet{poole08} and \citet[][also see \url{https://www.swift.ac.uk/analysis/uvot/}]{breeveld10}. No host-galaxy flux is subtracted from the \emph{Swift} photometry because we lack pre-explosion and late-time imaging but this is mitigated by the faintness of the host. 

LBT/MODS photometry was obtained in the $ugri$ filters which are analogous to the Sloan filter set. Images were calibrated with typical procedures including bias subtraction, dark subtraction, and flat-field corrections. The WCS solution was improved with the \textsc{astrometry.net} software \citep{lang2010} and the photometry was calibrated using the \textsc{refcat} catalog \citep{tonry2018b}. No host-galaxy subtraction is applied.

\subsection{Spectroscopy}

\begin{table}
    \centering
    \begin{tabular}{lrrrrl }
        Tel./Instr. & MJD & Phase [d] & Range & $\lambda/\delta \lambda$ & Notes \\
        \hline\hline
        UH2.2m/SNIFS & 60226.60 & 3.5 & $3400-9000$~\AAA & 1200 & \\
        NOT/AFOSC & 60230.25 & 7.1 & $3700-8800$~\AAA & 360 & Grism~4, 1\arcsec slit \\
        UH2.2m/SNIFS & 60231.59 & 8.5 & $3400-9000$~\AAA & 1200 & \\
        NOT/AFOSC & 60232.16 & 9.0 & $3700-8800$~\AAA & 360 & Grism~4, 1\arcsec slit \\
        NOT/AFOSC & 60236.18 & 13.0 & $3800-8200$~\AAA & 360 & Grism~4, 1\arcsec slit \\
        UH2.2m/SNIFS & 60237.60 & 14.4 & $3400-9000$~\AAA & 1200 & \\
        LBT/MODS & 60239.50 & 16.3 & $3100-10000$~\AAA & 2000 & \\
        Keck~II/KCWI & 60242.59 & 19.3 & $3500-8700$~\AAA & $1800,1000$ & BL+RL gratings \\
        UH2.2m/SNIFS & 60243.59 & 20.3 & $3400-9000$~\AAA & 1200 & \\
        UH2.2m/SNIFS & 60246.54 & 23.2 & $3400-9000$~\AAA & 1200 & \\
        UH2.2m/SNIFS & 60248.61 & 25.2 & $3400-9000$~\AAA & 1200 & \\
        LBT/MODS & 60254.45 & 31.0 & $3100-10000$~\AAA & 2000 & \\
        UH2.2m/SNIFS & 60271.48 & 47.8 & $3400-9000$~\AAA & 1200 & \\
        IRTF/SpeX & 60274.49 & 50.7 & $0.7-2.5$~\um & 100 & \\
        UH2.2m/SNIFS & 60274.57 & 50.8 & $3400-9000$~\AAA & 1200 & \\
        ANU2.3m/WiFeS & 60283.69 & 59.8 & $4500-8500$~\AAA & 3000 & B+R3000 gratings \\
        LBT/MODS & 60288.41 & 64.4 & $3100-10000$~\AAA & 2000 & \\
        ANU2.3m/WiFeS & 60291.68 & 67.7 & $4500-8500$~\AAA & 3000 & B+R3000 gratings \\
        Keck~II/KCWI & 60297.51 & 73.4 & $3500-8800$~\AAA & $1800,1000$ & BL+RL gratings \\
        Gemini-North/GMOS & 60303.11 & 78.9 & $4700-8900$~\AAA & 2000 & $R400$ grating, $G5305$ filter \\
        LBT/MODS & 60322.27 & 97.8 & $3100-10000$~\AAA & 2000 & \\
        Keck~II/KCWI & 60328.47 & 103.9 & $3500-8700$~\AAA & $1800,1000$ & BL+RL gratings \\
        IRTF/SpeX & 60335.46 & 110.8 & $0.7-2.5$~\um & 100 & \\
        IRTF/SpeX & 60344.43 & 119.6 & $0.7-2.5$~\um & 100 & \\
        Keck~II/KCWI & 60352.31 & 127.4 & $3600-6800$~\AAA & $1800,6500$ & BL+RH2 gratings \\
    \hline\hline
    \end{tabular}
    \caption{Information for the spectroscopic observations. Phases are given in rest-frame days. Range gives the wavelength coverage of the spectrum in \AAA (optical) or \um (NIR), although see Fig.~\ref{fig:allspex} for gaps in wavelength coverage for specific spectra. The approximate spectral resolution ($\lambda/\Delta \lambda)$ is provided for each observation. KCWI observations have 2 entries for spectral range and resolution corresponding to the blue and red channels (see Appendix~\ref{app:obs}).}
    \label{tab:specinfo}
\end{table}

We obtained 24 epochs of optical spectroscopy (Fig.~\ref{fig:allspex}) and 3 epochs of NIR spectroscopy (Fig.~\ref{fig:irspex}) using a multitude of telescopes and instruments. We briefly describe the individual data processing procedures below. The full log of spectra is provided in Table~\ref{tab:specinfo}.

\textit{UH2.2m+SNIFS}: 8 spectra were obtained with the University of Hawai`i 2.2m (UH2.2m) telescope using the SuperNova Integral Field Spectrograph \citep[SNIFS; ][]{lantz2004} thought the Spectroscopic Classification of Astronomical Transients \citep[SCAT; ][]{tucker2022}. SNIFS covers the full optical range ($\approx 3500-9000~\AAA$) at modest resolution ($R\sim 1200$). Data reduction and calibration procedures are described by \citet{tucker2022}. The dichroic crossover region ($\approx 5000-5200~\AAA$) is affected by changes in humidity so nights with poor corrections have these regions masked. 

\textit{NOT+ALOFC}: The 3 spectra from the Alhambra Faint Object Spectrograph (ALFOSC) on the Nordic Optical Telescope (NOT) were all obtained with Grism~4 using a 1\arcsec slit and 600~s exposures with no blocking filter under programme P68-007. All spectra were reduced using the \texttt{PyNOT-redux} reduction pipeline.\footnote{\url{github.com/jkrogager/PyNOT/}}

\textit{LBT+MODS}: 3 spectra were obtained with the Multi-Object Double Spectrograph \citep[MODS; ][]{pogge2010} on the Large Binocular Telescope (LBT). The data were reduced with \textsc{PypeIt} \citep{pypeit_prochaska} producing 4 spectra per epoch, one blue and one red channel per instrument (MODS1/2). These were then combined to produce the final spectra using weighted averages and sigma-clipping to remove outliers. 

\textit{Keck~II+KCWI}: 4 epochs of spectroscopy were obtained with the Keck Cosmic Web Imager \citep[KCWI; ][]{kcwi_morrissey} on the Keck~II telescope. The 2D images were reduced using \textsc{PypeIt} \citep{pypeit_prochaska}, using typical procedures such as bias subtraction, flat-fielding, and wavelength calibration using the FeAr arc lamps. Datacubes were created from these 2D images using the subpixel method within \textsc{pypeit\_coadd\_datacube}. The individual cubes were flux calibrated using a standard star observed on the same night. KCWI is an IFU so we extract spectra from the cubes by fitting analytic profiles to each slice of the 3D cube. The 2D extraction assumes a circular Gaussian profile for the seeing and a spatially flat sky background for each slice in the 3D $(x,y,\lambda)$ datacube. After extracting the 3D cube into 1D spectra the individual exposures are combined using sigma-clipped weighted averages to produce the final spectra. 

The host galaxy becomes visible as \name fades. We simultaneously extracted host and source spectra by fitting 2 Gaussian profiles plus a spatially uniform sky background using the coordinate offsets. The shape parameters of the galaxy are determined by initially fitting a slice of the 3D cube centered on bright emission lines (typically \Ha for the red channel and \OIIlong for the blue). Then, the shape and orientation are held fixed when measuring the integrated flux of each component as a function of wavelength in the 3D cube. Finally, the individual 1D spectra are combined to produce the final spectra shown in Fig.~\ref{fig:hostspec}. 

\textit{ANU2.3m+WiFeS}: Four spectra were obtained using the Wide Field Spectrograph (WiFeS) mounted on the Australian National University 2.3 metre telescope (ANU2.3m) located at Siding
Spring Observatory (SSO) \citep{2007Ap&SS.310..255D, 2010Ap&SS.327..245D}. Each spectrum was taken in `Nod \& Shuffle' mode using the $R=3000$ grating to cover the full $3000-9000~\AAA$ wavelength range. Each spectrum was reduced using PyWiFeS \citep[][(version 0.7.4]{wifes3}, with sky subtraction done using a 2D sky spectrum that was taken during the observations. See \citep{2024arXiv240213484C} for more details.  To increase the signal to noise of these spectra, we merged these four spectra into two which are used in the analysis.

\textit{Gemini-N+GMOS}: One spectrum is from the Gemini Multi-Object Spectrograph \citep[GMOS; ][]{hook2004} on the Gemini-North (GN) telescope. The spectrum used the $R400$ grating, the $G5305$ order-blocking filter, a 0\farcs5-wide slit, and $2\times2$ binning. The observations were reduced and extracted with \textsc{PypeIt} \citep{pypeit_prochaska}.

\textit{IRTF+SpeX}: 3 epochs of NIR spectroscopy were obtained using SpeX \citep{rayner03} on the NASA Infrared Telescope Facility (IRTF). The spectra were obtained in Prism mode at a low spectral resolution of R $\sim 100$. We reduced these spectra with Spextool \citep{cushing04} using flat and arc lamps taken immediately after the science spectra and then flux-calibrated the extracted 1D spectra using nearby A0 telluric standard stars within a typical airmass difference of $\approx$0.01.

\section{Additional Figures and Tables}\label{app:extra}

\begin{figure}
    \centering
    \includegraphics[width=\linewidth]{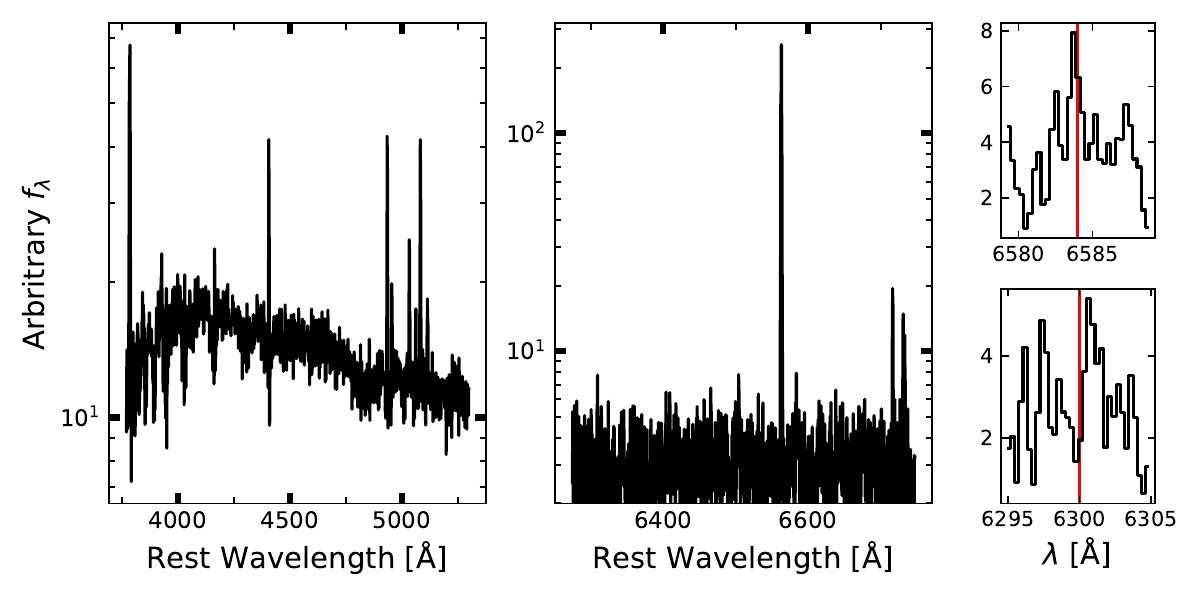}
    \caption{Host-galaxy spectrum extracted from the last KCWI observations. The small right panels shows insets around \NIIlong (top, marginally detected) and \OIa (undetected).}
    \label{fig:hostspec}
\end{figure}

\begin{figure}
    \centering
    \includegraphics[width=\linewidth]{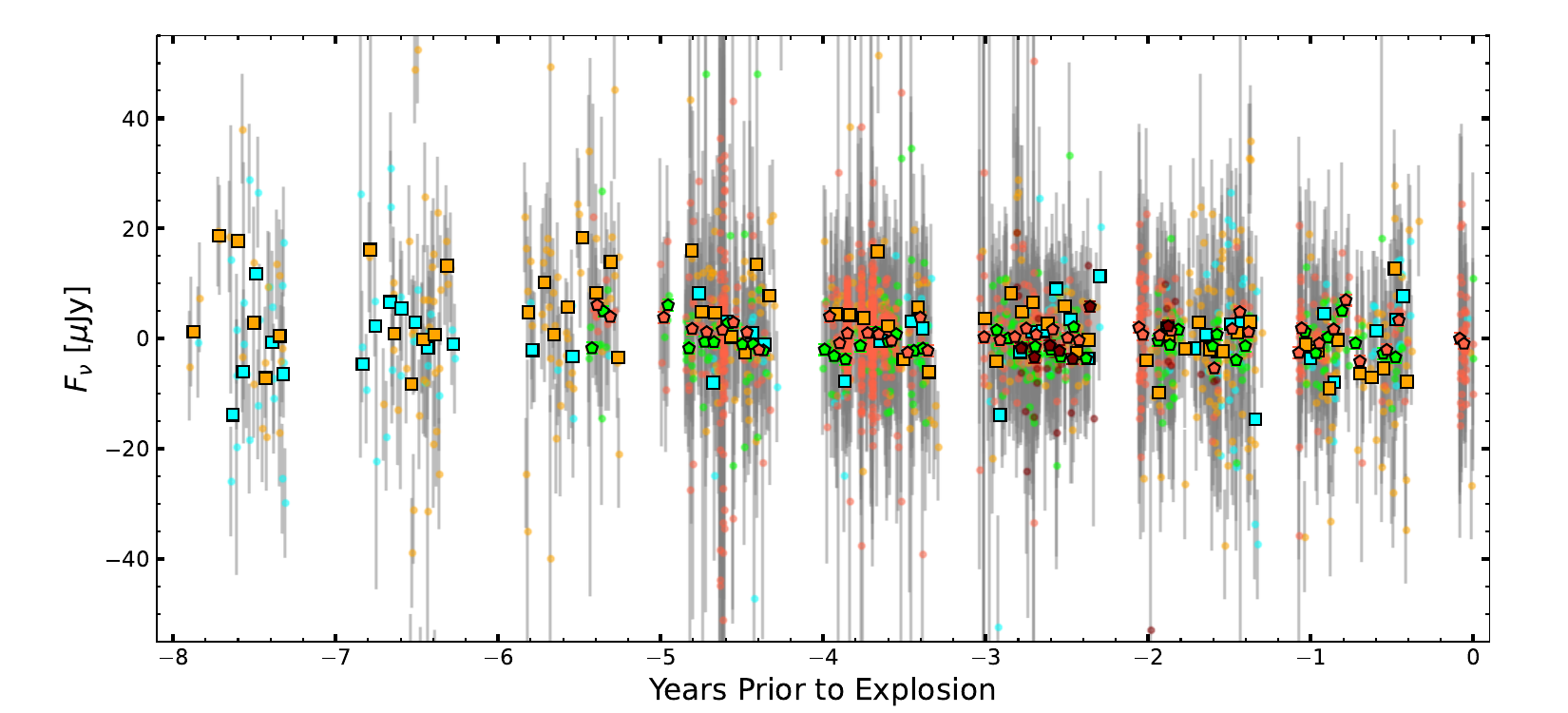}
    \caption{Pre-explosion survey photometry from ZTF and ATLAS showing no outbursts or variability in the 8 years preceding explosion. Single-night photometry is shown as small points with gray uncertainties whereas 30-day binned photometry is represented with larger bold points. Marker colors and symbols are the same as Fig.~\ref{fig:allphot}.
    }
    \label{fig:arxival}
\end{figure}

\begin{figure}
    \includegraphics[width=\linewidth]{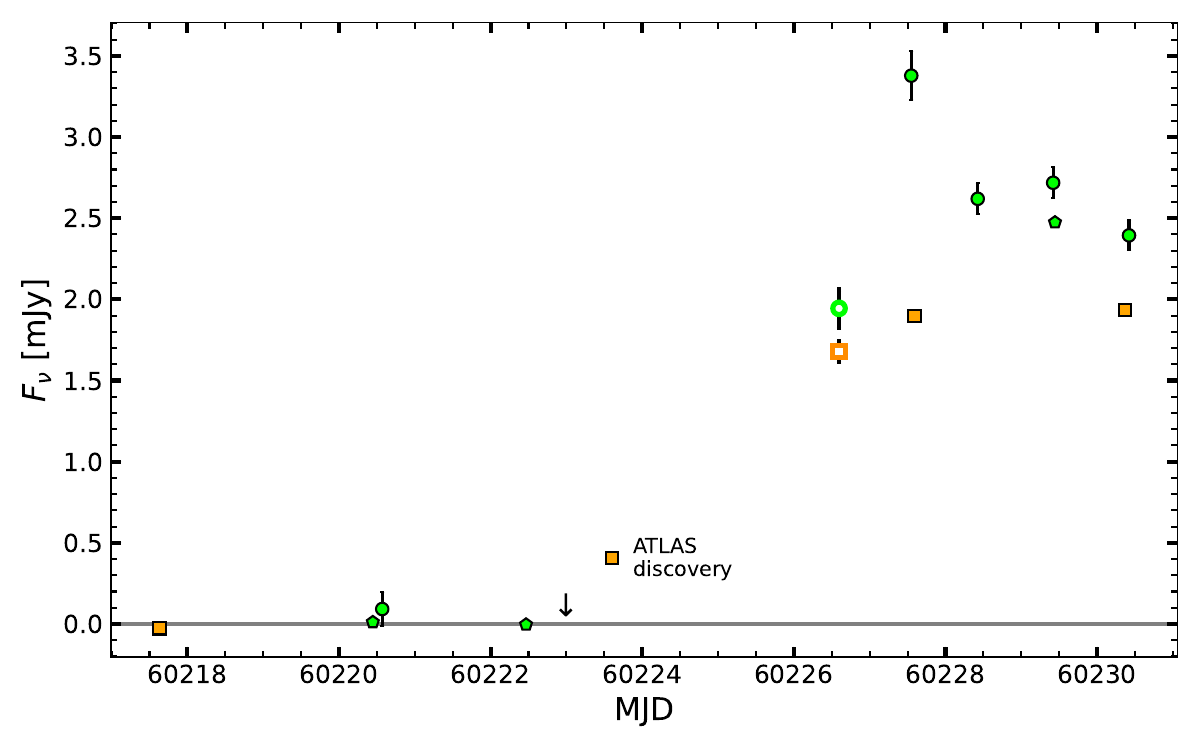}
    \caption{Early light curve of \name showing the adopted time of explosion (\texp) marked with a downwards arrow. Only $g$-band and $o$-band light curves are shown for clarity. Open symbols are synthesized from the first SNIFS spectrum.}
    \label{fig:earlyLC}
\end{figure}

\begin{figure}
    \centering
    \includegraphics{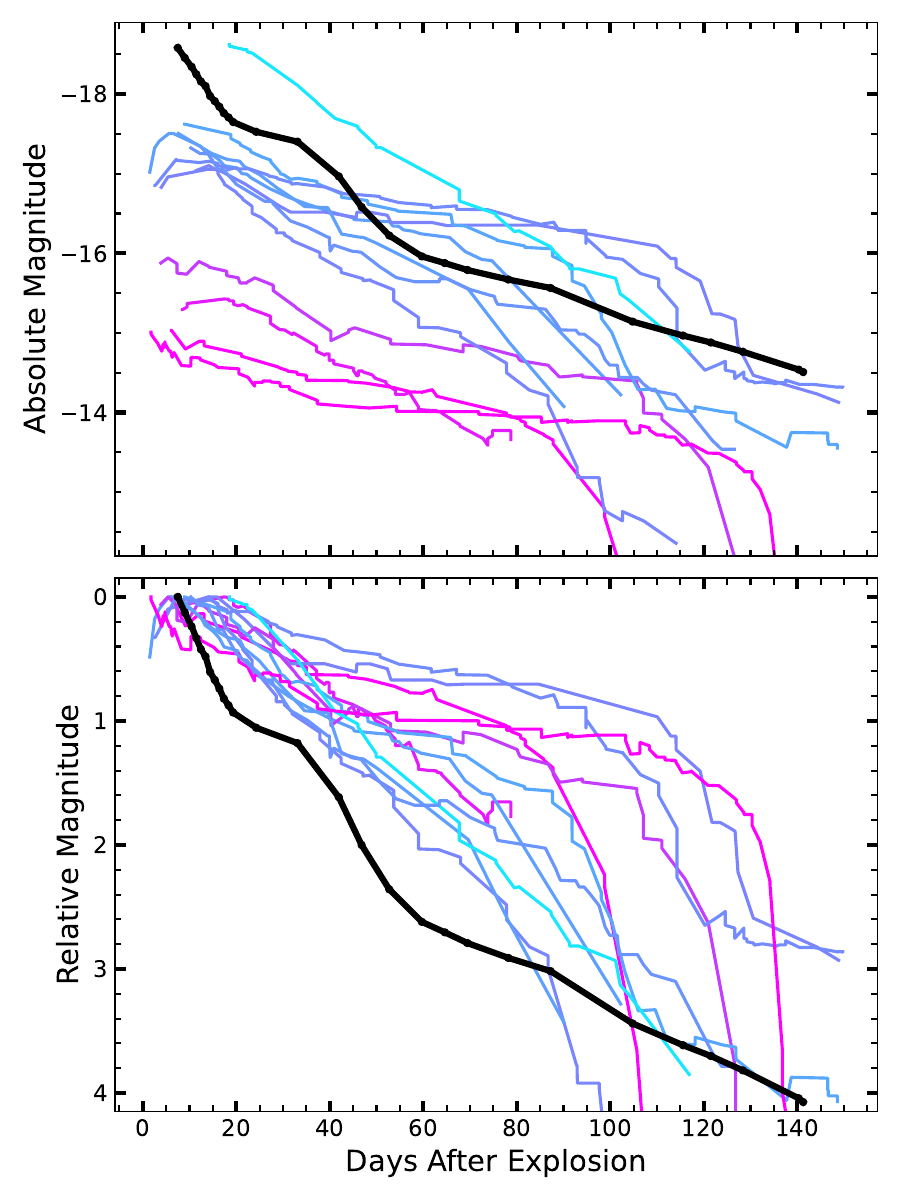}
    \caption{Comparing the $g$-band light curve of \name (black) to the sample from \citet{valenti2016}. We show the absolute magnitudes (top) and the magnitudes relative to peak (bottom) color-coded according to peak absolute magnitude.}
    \label{fig:gbandLC}
\end{figure}

Here we include additional figures and tables. Fig.~\ref{fig:hostspec} shows the extracted host-galaxy spectrum from the latest KCWI observations. Fig.~\ref{fig:arxival} shows the pre-explosion survey light curve. Fig.~\ref{fig:earlyLC} shows the photometric data used to infer \texp. Fig.~\ref{fig:gbandLC} compares the $g$-band light curve of name with the sample of \citet{valenti2016}.

Table~\ref{tab:reftable} includes references for the comparison \sneii used throughout the analysis. Comparison \sneii were chosen to include well-studied canonical SNe~IIP (SN~1999em, 2012aw, 2012ec, 2017eaw), SNe~IIL (2013ej, 2014G) and short-plateau \sneii (2006ai, 2006Y, 2016egz). SNe~2017gmr and 2017ivv were included because they share high absorption velocities and low host-galaxy metallicities, respectively, with \name. Most data are obtained from either the Weizmann Interactive Supernova Data Repository (WISeREP; \citealp{wiserep}) or the Open Supernova Catalog \citep[OSC; ][]{osc}.

The photometry (Fig.~\ref{fig:allphot}) and spectroscopy (Fig.~\ref{fig:allspex}) of \name are included in the online version of the manuscript. The measurements from Fig.~\ref{fig:Hevol} are also included as supplementary data.

\begin{table}[]
    \centering
    \begin{tabular}{l l}
    \hline
    Name & References \\\hline
    SN1999em &  \citet{1999em1}; \citet{1999em2}; \citet{1999em3}; \citet{1999em4}; \\
    & \citet{1999em5}; \citet{1999em6} \\
    SN2006ai &  \citet{hiramatsu2021} \\
    SN2006Y & \citet{hiramatsu2021} \\
    SN2012aw & \citet{2012aw1}; \citet{2012aw2}; \citet{2012aw3}; \citet{2012aw4}; \citet{2012aw5} \\
    SN2012ec & \citet{2012ec1}; \citet{2012ec2}; \citet{pessto}; \citet{awsnap} \\
    SN2013ej & \citet{2013ej1}; \citet{2013ej2}; \citet{bose2015}; \citet{2013ej4}; \citet{2013ej5}; \\ & \citet{yuan2016}\\
    SN2014G & \citet{2014G1}; \citet{2014G2} \\
    SN2016egz & \citet{hiramatsu2021}\\
%    SN2017eaw & \citet{2017eaw0}; \citet{2017eaw1}; \citet{2017eaw2}; \citet{2017eaw3}; \citet{2017eaw4} \\ 
    SN2017gmr & \citet{andrews2019} \\
    SN2017ivv & \citet{gutierrez2020} \\
    \hline
    \end{tabular}
    \caption{References for comparison SNe~II.}
    \label{tab:reftable}
\end{table}


\pagebreak
\bibliography{ref}{}
\bibliographystyle{aasjournal}



\end{document}